\pdfoutput=1
\documentclass[a4paper,11pt]{article}
\usepackage{amsmath,amssymb,amsfonts}
\usepackage[multiple,stable]{footmisc}
\usepackage{geometry}
\usepackage[amsmath,amsthm,thmmarks]{ntheorem}
\allowdisplaybreaks[4]
\usepackage[dvipsnames]{xcolor}
\usepackage[numbers,sort&compress]{natbib}
\usepackage[bf]{caption}
\usepackage{graphicx, subfig}
\usepackage{float}
\usepackage{tensor}
\usepackage[titles,subfigure]{tocloft}

\definecolor{bblue}{RGB}{18,107,174}
\usepackage[colorlinks=true, hyperfootnotes=true, linkcolor=blue, citecolor=bblue, anchorcolor=blue, urlcolor=magenta, linktocpage=true]{hyperref}

\newcommand{\beq}{\begin{equation}}
\newcommand{\eeq}{\end{equation}}
\newcommand{\bes}{\begin{equation*}}
\newcommand{\ees}{\end{equation*}}
\newcommand{\bea}{\begin{equation}\begin{aligned}}	
\newcommand{\eea}{\end{aligned}\end{equation}}	
\newcommand{\beas}{\begin{equation*}\begin{aligned}}  
\newcommand{\eeas}{\end{aligned}\end{equation*}}

\providecommand{\id}{\leavevmode\hbox{\small$\mathrm{1}$\kern-3.8pt\normalsize$\mathrm{1}$}}

\renewcommand{\thefootnote}{\fnsymbol{footnote}}
\begin{document}

\vspace{1cm}

\title{       {\Large \bf Soft gravitational radiation from multi-body collisions}}

\vspace{2cm}

\author{
Andrea Addazi$^{1,2}$\footnote{\url{addazi@scu.edu.cn}}\,\, and Kaiqiang Alan Zeng$^{1}$\footnote{\url{kaiqiangalanzeng@hotmail.com}}
}
\date{}
\maketitle
\begin{center} {\small $^{1}${\it Center for Theoretical Physics, College of Physics, Sichuan University, 610065 Chengdu, China} \\
$^{2}${\it Laboratori Nazionali di Frascati INFN Via Enrico Fermi 54, I-00044 Frascati (Roma), Italy}}

\end{center}

\rule[-8pt]{440pt}{0.08em}

\begin{abstract}
\noindent

We derive a universal expression for the gravitational radiation energy spectrum $dE^{GW}/d\omega$ at sub-leading order 
emitted from a generic gravitational hard scattering of multi-particles or multi-bodies. 
Our result includes all $\mathcal{O}(\omega)$ corrections
to the gravitational radiation flux from a generic 
$2\rightarrow N$ collision, in both the cases
of massless and massive particles/bodies. 
We also show  
the dependence of the radiation energy flux
 by the 
quantum spin in case of particle collisions. 
Then, we consider the specific case
of a gravitational elastic scattering of two massive bodies, 
i.e. $m+M\rightarrow m+M$ with $m,M$ the masses of the two bodies respectively.
We demonstrate that in this case 
all $\mathcal{O}(\omega)$ contributions 
to the energy flux 
exactly cancel each others. 
Nevertheless, we also show that, for a
 $2\rightarrow 2$ inelastic scattering,
the inclusion of sub-leading soft gravitons 
leads to a 
not zero radiation flux, having a simple expression in certain 
asymptotic regimes. 
Our results can be applied to the case of 
Black Hole collisions 
with possible testable implications 
in gravitational waves physics.

\end{abstract}

\thispagestyle{empty}

\rule[-8pt]{440pt}{0.08em}

\tableofcontents

\rule{440pt}{0.08em}

\setcounter{page}{1}

\renewcommand{\thefootnote}{\arabic{footnote}}

\section{Introduction}
\label{sec:1}

As it is well known, 
the gravitational bremsstrahlung radiation, emitted from a generic gravitational scattering, has a simple general 
expression in the leading-order (LO) zero-frequency limit (ZFL)
 derived many decades 
ago \cite{A1,A2,A3,A4,A5,A6,A7,A8}.
Nevertheless in the recent years
the interest 
to soft theorems has come back 
with several new results about sub- and sub-sub-leading 
terms beyond the  soft leading expression  
\cite{A9,A10,A11,A12,A13,A14,A15,A16}.
Indeed, many recent works were dedicated on several possible 
preservation and violation cases
of universality (of soft theorems)  \cite{A17,A18,A19,A20,A21,A22,A23,A24,A25,A26}
as well as to the connection of soft theorems
with the Bondi-Metzner-Van der Burg-Sachs (BMS)
symmetry group of asymptotically-flat space-time metric
\cite{A27,A28,A29,A30,A31,A32}.

Certainly, nowadays these theoretical issues,
that may appear in Weinberg's paper as academic, 
are now revitalized by the direct observations of 
Gravitational Waves from Black Hole (BH) and Neutron star
mergings \cite{LIGO}.
Indeed recently 
it was suggested that 
the {\it gravitational memory effect}
related to soft theorems and BMS 
may be tested from future gravitational waves experiments
\cite{Johnson:2019ljv}.
On the other hand, the analysis of gravitational radiation in soft regime 
can also be important for the detection of 
quantum gravity effects such possible $\alpha'$ corrections 
on radiation energy flux predicted by 
string theory in BH mergings
dubbed {\it string memory effect}
\cite{Addazi:2020obs,Aldi:2020qfu,Aldi:2021zhh,Ikeda:2021uvc}.

For these reasons, it is interesting to 
explore the possibility of 
using new tools developed 
from soft theorems and scattering amplitudes 
to extend the $dE^{GW}/d\omega$ 
results beyond the leading ZFL order, i.e. to the Next-to-leading order (NLO). 
Recent progresses of this program 
have been shown with different 
approaches, but substantial agreements of results
\cite{A38,A39,ABV,Sahoo&Sen,GSeikonal}.

On the other hand, ZFL radiation analysis can be 
compared with other results from the prospective of bremsstrahlung emission 
in high energy gravitational scatterings. 
In particular, a old standing program still in progress
explores aspects of gravitational scattering at the transplanckian energy limit 
\cite{40,41,42,43,44,45,46,55,56,57,58}
(e.g. see also \cite{59,60,61} for recent updates in the subject).
This can lead to important insights on the information paradox
in energy regimes where the BH 
should be generated \cite{47,48,49,50}
and should evaporate in form of Hawking radiation 
\cite{51}.
On the other hand, this
can also scrutinize string theory predictions 
from finite size of scattering vertices 
and tidal excitations \cite{40,43,52} 
as well as possible modifications of gravity at short-distance
and generalized uncertainty principle (GUP)
\cite{53,54}.


Let us consider a generic process from ${\bf IN}\rightarrow {\bf OUT}$, 
including virtual and real soft graviton effects. 
Let us denote the S-matrix of such a process as \footnote{
Let us also remark that in our paper we 
will focus on the $3+1$ space-time dimensional case.}
\beq
\label{SS}
\mathcal{S}^{(0)}\rightarrow \mathcal{S}={\rm exp}\Big(\frac{d^{3}q}{\sqrt{2\omega}}(\lambda_{q}^{*} a_{q}^{\dagger}-\lambda_{q} a_{q}) \Big)\mathcal{S}^{(0)}\, ,
\eeq
where $a_{q},a_{q}^{\dagger}$ are destruction and creation operators for  soft gravitons
of momenta $q$ and maximal cutoff energy $\Lambda<<E$ -- with $E$ the characteristic energy scale 
of the collision process -- and $\lambda_{q}$ is a process-dependent function of $q$.
Such an exponential operator applied on the bare S-matrix $\mathcal{S}^{0}$ corresponds to a coherent state operator
and soft gravitons are in a coherent vacuum state.

The energy spectrum of soft graviton emitted by a generic ${\bf IN}\rightarrow {\bf OUT}$ process 
is related to the eq. (\ref{SS}) as follows 
\beq
\frac{dE}{d^{3}q}^{GW}=\frac{\hbar}{2}|\lambda_{q}|^{2}\, , 
\eeq
where $\lambda_{q}$ depends from the specific ${\bf IN}\rightarrow {\bf OUT}$ scattering. 

In general such an expression can be expanded in powers of $q$
and, integrating on the momenta solid angle, 
one can obtain a $dE^{GW}/d\omega$ with a leading term
going in the ZFL $\omega\rightarrow 0$ as 
$dE_{0}^{GW}/d\omega\sim \omega^{-2}$.

Nevertheless,  a universal factorized expression for the sub-leading graviton emission 
is not known and it takes the form of a differential operator acting on the bare amplitudes. 
Indeed the subleading soft current depends by the total angular momentum operators
acting on the bare S-matrix. 

The main purpose of this paper is to compute the sub-leading order of the $dE^{GW}/d\omega$
spectrum. A first step towards it was done by {\it Bianchi}, {\it Veneziano} and one of the author 
of this paper \cite{ABV}, considering high energy ultra-relativistic scattering of spin-less particles.
Our aim is to extend the previous Addazi-Bianchi-Veneziano (ABV) analysis to the case of generic scatterings of massive 
particles and bodies, including non-relativistic regimes and having in mind 
the possible application of it for soft GW emission from BH mergings. 

As a warm up let us consider the first three leading orders 
of soft graviton emissions with universal behavior
\bea
\mathcal{M}_{N+1}\left(p_{i} ; q\right) =& \sqrt{8\pi G} \sum_{i=1}^{N}\left[\frac{p_{i} h p_{i}}{q p_{i}}+\frac{p_{i} h J_{i} q}{q p_{i}}-\frac{q J_{i} h J_{i} q}{2 q p_{i}}\right] \mathcal{M}_{N}\left(p_{i}\right)+\mathcal{O}(q^{2}) \\
\equiv& (S_{0}+S_{1}+S_{2})\mathcal{M}_{N}(p_{i})\, 
\label{eq:1-3}
\eea
where $\mathcal{M}_{N+1}\equiv \mathcal{M}_{N+1}(p_{i};q)$ denotes a generic $(N+1)$-particle on-shell scattering amplitude including an external graviton with momentum $q^{\mu}$ and polarization $h_{\mu\nu}$.

The $q^{\mu}$ and $h_{\mu\nu}$, defined above, satisfy 
\begin{alignat}{3} 
q^{2}=&0 \, ,&\qquad h_{\mu\nu}=&h_{\nu\mu} \,, &\qquad q^{\mu}h_{\mu\nu}&=0\, .
\label{eq:1-2}
\end{alignat}

The leading, subleading and sub-subleading soft factors are given by
\begin{alignat}{5}
S_{0}\equiv \sqrt{8\pi G} \sum_{i=1}^{N} \frac{p_{i} h p_{i}}{q p_{i}} \, , & \qquad & S_{1}=\sqrt{8\pi G} \sum_{i=1}^{N}\frac{p_{i}hJ_{i}q}{qp_{i}} \, , & \qquad & S_{2}=-\sqrt{8\pi G}\sum_{i=1}^{N}\frac{qJ_{i}hJ_{i}q}{2qp_{i}}
\label{eq:1-4}
\end{alignat}
and $J_{i}^{\mu\nu}\equiv p_{i}^{\mu}\partial_{i}^{\nu}-p_{i}^{\nu}\partial_{i}^{\mu}+S_{i}^{\mu\nu}$ denotes the total angular momentum of the $i$-th `hard' particle. 

The expression eq. (\ref{eq:1-3}) is gauge invariant under the rank-2 tensor shift
\beq
h_{\mu\nu}\rightarrow h_{\mu\nu}+q_{\mu}\zeta_{\nu}+q_{\nu}\zeta_{\mu}\, ,
\eeq
following from the conservation of momentum and angular momentum. 

The three soft terms can be related by the respective first three terms of the 
q-expansion of $\lambda_{q}$. 
This allows to compute the gravitational energy spectrum $dE^{GW}/d\omega$
after the integration over the emission direction. 
As we will see, such an operation involves not-trivial vector and tensor integrals 
that we perform in full generality. 

The plan of the paper is the following. 
In section \ref{sec:2} we will briefly review Weinberg's derivation of the 
leading $B$-factor and $dE^{GW}/d\omega$. 
In section \ref{sec:3}, we will show a complete computation of the sub-leading emission of soft radiation,
in the case of massive particle collisions, after a review of the previous ABV result. 
In section \ref{sec:4} \& \ref{sec:5}, we will show a new surprising result: in case of $2\rightarrow 2$ elastic scattering 
all sub-leading effects exactly vanish for generic particle/body collision. 
In section \ref{sec:6}, we analyze the case of $2\rightarrow 2$ inelastic collision: we find an exact non-zero result 
which is, in full generality, complicated, but it has simpler expressions in certain kinematic regimes. 

\section{Soft gravitational radiation from leading order: short review}
\label{sec:2}
The soft theorem up to leading order is universal and the dominant behavior reads
\beq
|\mathcal{M}_{N+1}(p_i;q)|^2=8\pi G\sum_{s={\pm 2}}\left|\sum^{N}_{i=1}\frac{p_{i}h^{s}p_{i}}{qp_i}\right|^2|\mathcal{M}_{N}(p_i)|^2 \, ,
\label{eq:2-1}
\eeq
where notation is defined in the introduction above. 
The polarization tensor $h^{s}$ satisfies $h_{\mu\nu}^{-s}=(h^s_{\mu\nu})^{*}$ and after tedious but straightforward manipulations one can obtain that 
the sum on the spins of the $hh$ 4-tensor corresponds to 
\beq
\sum_{s=\pm 2} h^{s}_{\mu\nu}h^{-s}_{\rho\sigma}=\Pi_{\mu\nu,\rho\sigma}=\frac{1}{2}(\pi_{\mu\rho}\pi_{\nu\sigma}+\pi_{\mu\sigma}\pi_{\nu\rho}-\pi_{\mu\nu}\pi_{\rho\sigma})
\, , 
\label{eq:2-2}
\eeq
where
\beq
\pi_{\mu\nu}=\eta_{\mu\nu}-q_{\mu}\bar{q}_{\nu}-q_{\nu}\bar{q}_{\mu}\, ,\quad \bar{q}^2=0\quad \textrm{and} \quad \bar{q}q=1\, .
\label{eq:2-3}
\eeq

Weinberg's B-factor can be obtained after the 
 integration over the momentum of the soft graviton in the final state $q=|q|(-1,\vec{n})$:

\bes
B_{0}= \int \frac{d^{3}q}{2|q|(2\pi)^{3}}|\mathcal{M}_{N+1}|^{2} 
\ees
which corresponds to the number density as follows
\beq
\frac{dN_{0}}{d\omega}=\frac{dB_{0}}{d\omega}= \int \frac{d^{3}q \delta(|q|-\omega)}{2|q|(2\pi)^{3}}|\mathcal{M}_{N+1}|^{2}\,
\label{eq:2-4}
\eeq 

and the energy spectrum of the gravitational wave is then
\beq 
\frac{dE_{0}^{\textrm{GW}}}{d\omega}=\hbar \omega \frac{dN_{0}}{d\omega}\, .
\label{eq:2-5}
\eeq

To evaluate the soft factor in eq. (\ref{eq:2-1}), one first needs to compute the polarization sum and the result reads as 
\beq
\sum_{i,j}\frac{p_{i}^{\mu}p_{i}^{\nu}}{qp_{i}}\Pi_{\mu\nu,\rho\sigma}\frac{p_{j}^{\rho}p_{j}^{\sigma}}{qp_{j}}= \sum_{i,j}\frac{\left(p_{i}p_{j}\right)^{2}-\frac{1}{2}p_{i}^{2}p_{j}^{2}}{qp_{i}qp_{j}}= \sum_{i,j}\frac{\left(p_{i}p_{j}\right)^{2}-\frac{1}{2}m_{i}^{2}m_{j}^{2}}{qp_{i}qp_{j}}\, .
\label{eq:2-1-1}
\eeq

The evaluation of the integral in the $B$-factor 
\beq
I= \int \frac{d^{3}q}{|q|qp_{i}qp_{j}}
\label{eq:2-1-2}
\eeq
leads to the well know result 
\beq
I=-\ln \frac{\Lambda}{\lambda} \left(\frac{2\pi}{\beta_{ij}p_{i}p_{j}}\right) \ln\frac{1+\beta_{ij}}{1-\beta_{ij}}\, ,
\label{eq:2-1-3}
\eeq
where we have used $\lambda$ and $\Lambda$ to denote the IR cutoff scale and the upper limit for the validity of the leading soft behavior respectively. Finally, one obtains the well-known 
Weinberg B-factor \cite{A2,2,ABV}: 
\bea
B_{0}=& -\frac{8\pi G}{2(2\pi)^3}\sum_{i,j}\frac{(p_{i}p_{j})^2-\frac{1}{2}m_{i}^{2}m_{j}^{2}}{E_{i}E_{j}}\ln \frac{\Lambda}{\lambda}\frac{2\pi E_{i}E_{j}}{\beta_{ij}p_{i}p_{j}} \ln \frac{1+\beta_{ij}}{1-\beta_{ij}}\\
=& -\frac{G}{2\pi} \ln \frac{\Lambda}{\lambda} \sum_{i,j}\frac{(p_{i}p_{j})(1+\beta_{ij}^{2})}{\beta_{ij}} \ln \frac{1+\beta_{ij}}{1-\beta_{ij}}\, .
\label{eq:2-1-4}
\eea

\section{Soft gravitational radiation from sub-leading order}
\label{sec:3}
The sub-leading contribution to the GW spectrum comes from the interference between $S_{0}$ and $S_{1}$ of eq. (\ref{eq:1-3}) 
\beq
B_{1}=8 \pi G \int \frac{d^{3} q}{2|q|(2 \pi)^{3}} \sum_{i, j} \sum_{s=\pm 2}\left[\frac{\left(p_{i} h^{s} p_{i}\right)\left(p_{j} h^{(-s)} J_{j} q\right)}{q p_{i} q p_{j}}+(i \leftrightarrow j)\right]\, .
\label{eq:3-1}
\eeq

Now, contrary to Winberg's $B_{0}$, the sub-leading factor $B_{1}$ involves the presence of the angular momentum operator though as acting on the $S$-matrix. 
\subsection{Massless Case: a review of the ABV result}
\label{sec:3.1}
In this section, we will review the previous ABV result as an introduction to the main results of our paper.

Let us consider the case in which the `hard' particles are massless:
the sum over helicities of the emitted soft gravitons is 
\bea
\sum_{i,j}\frac{p_{i}^{\mu}p_{i}^{\nu}p_{j}^{\rho}\left(J_{j}q\right)^{\sigma}}{qp_{i}qp_{j}}\Pi_{\mu\nu,\rho\sigma}+(i\leftrightarrow j)=& \sum_{i,j}\frac{\left(p_{i}p_{j}\right)\left(p_{i}J_{j}q\right)}{qp_{i}qp_{j}}+(i\leftrightarrow j) \\
=& \sum_{i, j} \frac{p_{i} p_{j}}{q p_{i} q p_{j}}\left[p_{i} J_{j} q+p_{j} J_{i} q\right] \, .
\label{eq:3-1-1}
\eea

Thus, $B_{1}$ factor can be rewritten in a simplified form as 
\beq
B_{1}=8 \pi G \int \frac{d^{3} q}{2|q|(2 \pi)^{3}} \sum_{i, j} \frac{p_{i} p_{j}}{q p_{i} q p_{j}}\left[p_{i} \overrightarrow{J}_{j}+p_{j} \overleftarrow{J}_{i}\right] q \, , 
\label{eq:3-1-2}
\eeq
where $\overleftarrow{J}_{i}$ and $\overrightarrow{J}_{j}$ act on $\mathcal{M}_{N}$ and its complex conjugate $\mathcal{M}_{N}^{*}$ respectively.

The four-vector integral that we need to evaluate for our purposes is 
\beq
I_{i j}^{\mu}=\int \frac{d^{3} q}{2|q|(2 \pi)^{3}} \frac{p_{i} p_{j} q^{\mu}}{q p_{i} q p_{j}}=\int \frac{d^{4} q}{(2 \pi)^{3}} \delta_{+}\left(q^{2}\right) \frac{p_{i} p_{j} q^{\mu}}{q p_{i} q p_{j}} \, ,
\label{eq:3-1-3}
\eeq
where we have defined $\delta_{+} (q^{2})\equiv \delta(q^{2})\Theta(-q_{0})$ and we used 
\beq
\int^{\infty}_{-\infty} dq_{0} \delta(q^{2})\Theta(-q_{0})= \frac{1}{2|q|}\, , 
\label{eq:3-1-4}
\eeq
where the Heaviside step function constrains the energy to be positive.

When one considers the angel between $\vec{q}$ and $\vec{p}_{i}$ going to zero, 
apparent collinear divergences appear in the integral eq. (\ref{eq:3-1-3}). 
This can be explicitly seen as $qp_{i}=|q|E_{i}(-1+\cos \theta_{qi})\sim |q|E_{i}\theta^{2}_{qi}$ as $\theta_{qi}\rightarrow 0$ while $\sin \theta_{qi} d\theta_{qi}$ inside the phase space integral $d^{3}q$ goes like $\theta_{qi}d\theta_{qi}$, thus the $\theta_{qi}$-integral becomes $d \ln \theta_{qi}$ which is divergent for $\theta_{qi}\rightarrow 0$ as well as for $\theta_{qj}$. 
However, we would expect that gravity is not an interaction plagued by collinear divergences. 
Thus, in order to avoid these apparent divergences, ABV \cite{ABV} introduced the following $ij$-sum zero-equivalent shift terms as follows 
\beq
I_{i j}^{\mu} \rightarrow \tilde{I}_{i j}^{\mu}=\int \frac{d^{4} q}{(2 \pi)^{3}} \delta_{+}(q^{2}) \frac{[(p_{i} p_{j}) q^{\mu}-(q p_{j}) p_{i}^{\mu}-(q p_{i}) p_{j}^{\mu}]}{(p_{i} q)(p_{j} q)}\, ,
\label{eq:3-1-5}
\eeq
the shift terms in eq. (\ref{eq:3-1-5}) vanish after summing over $i$ and $j$ thanks to momentum and angular momentum conservations. 

The four-vector integral eq. (\ref{eq:3-1-5}) can be rewritten in a  Lorentz contravariant form as follows 
\beq
K^{\mu}_{ij}(P,\Lambda)=\int \frac{d^{4}q}{(2\pi)^{3}}\frac{\delta_{+}(q^{2})\delta \left(\frac{qP}{\Lambda^{2}}+1\right)}{(qp_{i})(qp_{j})}[(p_{i}p_{j})q-(qp_{j})p_{i}-(qp_{i})p_{j}]^{\mu}\, ,
\label{eq:3-1-6}
\eeq 
where $P$ is an arbitrary four vector and $\Lambda$ is a constant with the dimension of energy. 

In a generic $n\rightarrow m$ process, $P$ can be identified as the total momentum of $n$ incoming (or $m$ outgoing) particles
and thus we can choose it as $P=(E_{\mathrm{CM}},0)$. 
With this choice, $\Lambda^{2}=\sqrt{s}\hbar\omega_{0}$ where $s=-P^{2}=E_{\mathrm{CM}}^{2}$ is the Mandelstam variable of the corresponding channel and $\omega_{0}$ is the center-of-mass frequency at which we wish to compute the energy spectrum.
Then the Dirac-Delta can be rewritten as 
\bes
\delta\left(\frac{qP}{\Lambda^{2}}+1\right)=\delta \left(1-\frac{\omega}{\omega_{0}}\right)\, .
\ees

The Lorentz-invariant (graviton number) spectrum ${dB_{1}}/{d\omega_{0}}$ is then given by
\beq
\frac{dB_{1}}{d\omega_{0}}= 8\pi G \sum_{i,j} \frac{K^{\mu}_{ij}(P,\omega_{0})}{\omega_{0}}[p_{i}\overrightarrow{J}_{j}+p_{j}\overleftarrow{J}_{i}]_{\mu}\, .
\label{eq:3-1-7}
\eeq

With Lorentz covariance, $K_{ij}^{\mu}$ can be expanded as 
\beq
K^{\mu}_{ij}(p_{i}p_{j})= K_{P}(s)P^{\mu}+K_{i}(Pp_{i})p^{\mu}_{i}+K_{j}(Pp_{j})p^{\mu}_{j}\, ,
\label{eq:3-1-8}
\eeq
since $K_{ij}^{\mu}$ is orthogonal to both $p_{i}$ and $p_{j}$, $K^{\mu}_{ij}$ can be rewritten in the form
\beq
K^{\mu}_{ij}=K[(p_{i}p_{j})P^{\mu}-(Pp_{j})p_{i}^{\mu}-(Pp_{i})p_{j}^{\mu}]\equiv K(p_{i}p_{j})Q^{\mu}_{ij} \, , 
\label{eq:3-1-9}
\eeq
where we have defined $K\equiv \frac{K_{P}}{p_{i}p_{j}}$ 
and a new vector
\beq
Q^{\mu}_{ij}\equiv P^{\mu}-\frac{Pp_{j}}{p_{i}p_{j}}p_{i}^{\mu}-\frac{Pp_{i}}{p_{i}p_{j}}p_{j}^{\mu}\, .
\label{eq:3-1-10}
\eeq

The main integrals involved in eq. (\ref{eq:3-1-7}) are
\beq
\int\frac{d^{3}q}{|q|}\delta(\omega-\omega_{0})\frac{(p_{i}p_{j})(qP)}{qp_{i}qp_{j}}= -\frac{(p_{i}p_{j})\sqrt{s}}{E_{i}E_{j}}J= \frac{2\pi \sqrt{s}}{\beta_{ij}}\ln \frac{1+\beta_{ij}}{1-\beta_{ij}} \, ,
\label{eq:3-1-11}
\eeq
and
\bea
\int\frac{d^{3}q}{|q|}\delta(\omega-\omega_{0})\left(\frac{Pp_{i}}{qp_{i}}+\frac{Pp_{j}}{qp_{j}}\right) =&  \sqrt{s}\left(L_{i}+L_{j}\right) \\
=& -2\pi\sqrt{s}  \left(\frac{1}{|\vec{v}_{i}|}\ln\frac{1-|\vec{v}_{i}|}{1+|\vec{v}_{i}|}+\frac{1}{|\vec{v}_{j}|}\ln\frac{1-|\vec{v}_{j}|}{1+|\vec{v}_{j}|}\right)\, .
\label{eq:3-1-12}
\eea
 In the massless limit, the above expressions has the form
\beq
\int\frac{d^{3}q}{|q|}\delta(\omega-\omega_{0})\frac{(p_{i}p_{j})(qP)}{qp_{i}qp_{j}}= 4\pi\sqrt{s} \ln\left(-\frac{2p_{i}p_{j}}{m_{i}m_{j}}\right)\, ,
\label{eq:3-1-13}
\eeq
and 
\beq
\int\frac{d^{3}q}{|q|}\delta(\omega-\omega_{0})\left(\frac{Pp_{i}}{qp_{i}}+\frac{Pp_{j}}{qp_{j}}\right)= -4\pi\sqrt{s} \ln\frac{m_{i}m_{j}}{4E_{i}E_{j}}\, .
\label{eq:3-1-14}
\eeq

Contracting $K^{\mu}_{ij}$ with $P^{\mu}$, we get from eq. (\ref{eq:3-1-6}): 
\bea
K_{ij}P=& \int \frac{d^{3}q}{2|q|(2\pi)^{3}}\delta \left(1-\frac{\omega}{\omega_{0}}\right)\left[\frac{(p_{i}p_{j})(Pq)}{(qp_{i})(qp_{j})}-\left(\frac{Pp_{i}}{qp_{i}}+\frac{Pp_{j}}{qp_{j}}\right)\right]\\
=& -\frac{4\pi \omega_{0} \sqrt{s}}{2(2\pi)^{3}}\left[\ln \left(-\frac{2p_{i}p_{j}}{m_{i}m_{j}}\right) +\ln \left(\frac{m_{i}m_{j}}{4E_{i}E_{j}}\right) \right]
= -\frac{\omega_{0}\sqrt{s}}{4\pi^{2}}\ln \left(-\frac{p_{i}p_{j}}{2E_{i}E_{j}}\right)
\label{eq:3-1-15}
\eea
and from eq. (\ref{eq:3-1-9}):
\beq
K_{ij}P=K[(p_{i}p_{j})P^{2}-2(Pp_{j})(Pp_{i})]\, .
\label{eq:3-1-16}
\eeq

$K$ is then determined by equating eq. (\ref{eq:3-1-15}) and eq. (\ref{eq:3-1-16}) 
\beq
K(p_{i}p_{j})= \frac{\omega_{0}\sqrt{s}}{4\pi^{2}\tilde{s}_{ij}}\ln \left(-\frac{p_{i}p_{j}}{2E_{i}E_{j}}\right) \, ,
\label{eq:3-1-17}
\eeq
where we have introduced the quantity 
\beq
\tilde{s}_{ij}=-Q^{2}_{ij}=s+\frac{2(Pp_{i})(Pp_{j})}{p_{i}p_{j}} \, . 
\label{eq:3-1-18}
\eeq

Substituting the above result into eq. (\ref{eq:3-1-6}) and renaming $\omega_{0}$ as $\omega$, we have a sub-leading factor as follows:
\bea
\frac{dB_{1}}{d\omega}=& -2\frac{ G\sqrt{s}}{\pi} \sum_{i,j}\frac{1}{\tilde{s}_{ij}} \ln \left(-\frac{p_{i}p_{j}}{2E_{i}E_{j}}\right) Q^{\mu}_{ij}[p_{i}\overrightarrow{J}_{j}+p_{j}\overleftarrow{J}_{i}]_{\mu}\\
=& -2\frac{G\sqrt{s}}{\pi} \sum_{i,j}\frac{(p_{i}p_{j})}{\tilde{s}_{ij}} \ln \left(-\frac{p_{i}p_{j}}{2E_{i}E_{j}}\right) Q^{\mu}_{ij}[\overrightarrow{\partial}_{j}+\overleftarrow{\partial}_{i}]_{\mu} \, .
\label{eq:3-1-19}
\eea

\begin{figure}[tbp]
\centering
\includegraphics[width=.5\textwidth]{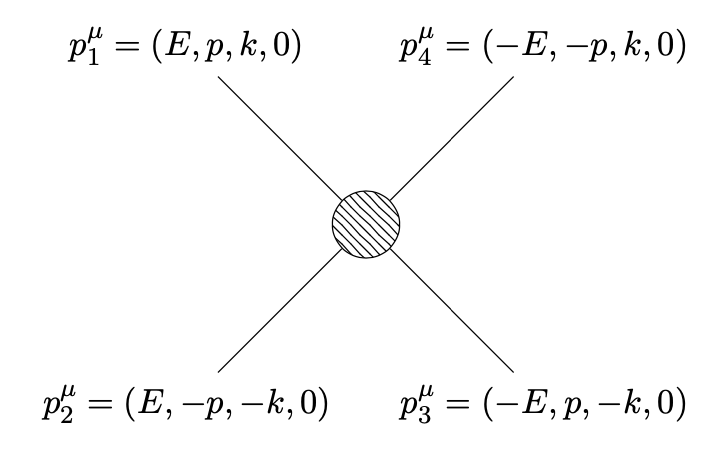}
 \caption{$2\rightarrow 2$ scattering in Breit frame.}
 \label{FIG:1}
\end{figure}

In \cite{ABV}, ABV showed that the sub-leading correction ${dB_{1}}/{d\omega}$ vanishes for a $2\rightarrow 2$ process by direct computation in the Briet frame (see Fig.\ref{FIG:1}). We will show in the following sections that the zero result is also re-obtained in the massive case. In next section, we will derive a new $B$-factor for the massive case.
\subsection{Massive case: a generalization of ABV result}
\label{sec:3.2}
In this section, we will generalize the above discussion to the massive case.

First, we notice that the polarization sum in the massive case has an additional term and it reads as 
\bea
&\sum_{i,j}\frac{p_{i}^{\mu}p_{i}^{\nu}p_{j}^{\rho}\left(J_{j}q\right)^{\sigma}}{qp_{i}qp_{j}}\Pi_{\mu\nu,\rho\sigma}+(i\leftrightarrow j)\\
= &\sum_{i,j}\frac{\left(p_{i}p_{j}\right)\left(p_{i}J_{j}q\right)-\frac{1}{2}p_{i}^{2}\left(p_{j}J_{j}q\right)}{qp_{i}qp_{j}}+(i\leftrightarrow j) \\
=& \sum_{i,j}\frac{p_{i}p_{j}}{qp_{i}qp_{j}}\left[p_{i}J_{j}q+p_{j}J_{i}q\right]-\frac{1}{2}\frac{1}{qp_{i}qp_{j}}\left[p_{i}^{2}\left(p_{j}J_{j}q\right)+p_{j}^{2}\left(p_{i}J_{i}q\right)\right] \, .
\label{eq:3-2-1}
\eea

Considering the polarization sum in eq. (\ref{eq:3-1}) with eq. (\ref{eq:3-2-1}), the sub-leading $B$-factor becomes
\bea
B_{1} =& 8\pi G\int\frac{d^{3}q}{2|q|(2\pi)^{3}}\sum_{i,j}\left\{\frac{p_{i}p_{j}q^{\mu}}{qp_{i}qp_{j}}\left(p_{i}\overrightarrow{J}_{j}+p_{j}\overleftarrow{J}_{i}\right)_{\mu}\right.\\
&\left. -\frac{1}{2}\frac{q^{\mu}}{qp_{i}qp_{j}}\left[p_{i}^{2}\left(p_{j}\overrightarrow{J}_{j}\right)_{\mu}+p_{j}^{2}\left(p_{i}\overleftarrow{J}_{i}\right)_{\mu}\right]\right\}\, .
\label{eq:3-2-2}
\eea

Then $B_{1}$ can be rewritten in terms of the integral eq. (\ref{eq:3-1-3}) as follows
\beq
B_{1}=8\pi G\sum_{i,j}I^{\mu}_{ij}\left\{\left(p_{i}\overrightarrow{J}_{j}+p_{j}\overleftarrow{J}_{i}\right)_{\mu}-\frac{1}{2}\left[\frac{p_{i}^{2}}{p_{i}p_{j}}\left(p_{j}\overrightarrow{J}_{j}\right)_{\mu}+\frac{p_{j}^{2}}{p_{i}p_{j}}\left(p_{i}\overleftarrow{J}_{i}\right)_{\mu}\right]\right\}\, .
\label{eq:3-2-3}
\eeq

Following the same procedure in section \ref{sec:3.1}, one gets the number density spectrum 
\beq
\frac{dB_{1}}{d\omega_{0}} = -8\pi G\sum_{i,j}\frac{K_{ij}^{\mu}(P,\omega_{0})}{\omega_{0}}\left[p_{i}\overrightarrow{J}_{j}-\frac{1}{2}\frac{p_{i}^{2}}{p_{i}p_{j}}\left(p_{j}\overrightarrow{J}_{j}\right)+(i\leftrightarrow j)\right]
\label{eq:3-2-4}
\eeq
where $K_{ij}^{\mu}$ is given by  
\beq
K_{ij}^{\mu}(P,\omega_{0})=\int\frac{d^{4}q}{(2\pi)^{3}}\frac{\delta_{+}\left(q^{2}\right)\delta\left(1-\frac{\omega}{\omega_{0}}\right)}{\left(qp_{i}\right)\left(qp_{j}\right)}\left[\left(p_{i}p_{j}\right)q-\left(qp_{j}\right)p_{i}-\left(qp_{i}\right)p_{j}\right]^{\mu}\, ,
\label{eq:3-2-5}
\eeq 
and it can be written in the form
\beq
K_{ij}^{\mu}= K\left[\left(p_{i}p_{j}\right)P^{\mu}-\left(Pp_{j}\right)p_{i}^{\mu}-\left(Pp_{i}\right)p_{j}^{\mu}\right]\equiv K\left(p_{i}p_{j}\right)Q_{ij}^{\mu}\, .
\label{eq:3-2-6}
\eeq

Contracting $K_{ij}^{\mu}$ with $P_{\mu}$, with eq. (\ref{eq:3-1-11}) and eq. (\ref{eq:3-1-13}) and from eq. (\ref{eq:3-2-5}), one obtains 
\beas
K_{ij}P=& \int\frac{d^{3}q}{2|q|(2\pi)^{3}}\delta\left(1-\frac{\omega}{\omega_{0}}\right)\left[\frac{\left(p_{i}p_{j}\right)(qP)}{qp_{i}qp_{j}}-\left(\frac{Pp_{i}}{qp_{i}}+\frac{Pp_{j}}{qp_{j}}\right)\right]\\
=& -\frac{\omega_{0}}{2(2\pi)^{3}}\left[\frac{2\pi\sqrt{s}}{\beta_{ij}}\ln\frac{1+\beta_{ij}}{1-\beta_{ij}}+2\pi\sqrt{s}\left(\frac{1}{\left|\vec{v}_{i}\right|}\ln\frac{1-\left|\vec{v}_{i}\right|}{1+\left|\vec{v}_{i}\right|}+\frac{1}{\left|\vec{v}_{j}\right|}\ln\frac{1-\left|\vec{v}_{j}\right|}{1+\left|\vec{v}_{j}\right|}\right)\right]\, , 
\eeas
and from eq. (\ref{eq:3-2-6}) 
\bes
K_{ij}P=K\left[\left(p_{i}p_{j}\right)P^{2}-2\left(Pp_{j}\right)\left(Pp_{i}\right)\right]\, .
\ees
Thus $K$ is determined by
\beq
K(p_{i}p_{j}) = \frac{\omega_{0}\sqrt{s}}{8\pi^{2}\tilde{s}_{ij}}\left[\frac{1}{\beta_{ij}}\ln\frac{1+\beta_{ij}}{1-\beta_{ij}}+\left(\frac{1}{\left|\vec{v}_{i}\right|}\ln\frac{1-\left|\vec{v}_{i}\right|}{1+\left|\vec{v}_{i}\right|}+\frac{1}{\left|\vec{v}_{j}\right|}\ln\frac{1-\left|\vec{v}_{j}\right|}{1+\left|\vec{v}_{j}\right|}\right)\right]\, .
\label{eq:3-2-7}
\eeq

Substituting the above $K$-factor into eq. (\ref{eq:3-2-4}) gives
\bea
\frac{dB_{1}}{d\omega}=& -\frac{G\sqrt{s}}{\pi}\sum_{i,j}\frac{1}{\tilde{s}_{ij}}\left[\frac{1}{\beta_{ij}}\ln\frac{1+\beta_{ij}}{1-\beta_{ij}}+\left(\frac{1}{\left|\vec{v}_{i}\right|}\ln\frac{1-\left|\vec{v}_{i}\right|}{1+\left|\vec{v}_{i}\right|}+\frac{1}{\left|\vec{v}_{j}\right|}\ln\frac{1-\left|\vec{v}_{j}\right|}{1+\left|\vec{v}_{j}\right|}\right)\right]\\
&\times Q_{ij}^{\mu}\left[p_{i}\overrightarrow{J}_{j}+p_{j}\overleftarrow{J}_{i}\right]_{\mu}\, . 
\label{eq:3-2-8}
\eea

Considering that
\bes
Q_{ij}^{\mu}\left[p_{i}\overrightarrow{J}_{j}+p_{j}\overleftarrow{J}_{i}\right]_{\mu}=(p_{i}p_{j})Q_{ij}^{\mu}\left(\overleftarrow{\partial}_{i}+\overrightarrow{\partial}_{j}\right)_{\mu}+\frac{Pp_{i}}{p_{i}p_{j}}p_{j}^{2}\left(\overrightarrow{\partial}_{j}p_{i}\right)+\frac{Pp_{j}}{p_{i}p_{j}}p_{i}^{2}\left(\overleftarrow{\partial}_{i}p_{j}\right)\, ,
\ees
we finally obtain the new general subleading $B$-factor for the massive case
\bea
\frac{dB_{1}}{d\omega}=& -\frac{G\sqrt{s}}{\pi}\sum_{i,j}\frac{1}{\tilde{s}_{ij}}\left[\frac{1}{\beta_{ij}}\ln\frac{1+\beta_{ij}}{1-\beta_{ij}}+\left(\frac{1}{\left|\vec{v}_{i}\right|}\ln\frac{1-\left|\vec{v}_{i}\right|}{1+\left|\vec{v}_{i}\right|}+\frac{1}{\left|\vec{v}_{j}\right|}\ln\frac{1-\left|\vec{v}_{j}\right|}{1+\left|\vec{v}_{j}\right|}\right)\right] \\
& \times\left[(p_{i}p_{j})Q_{ij}^{\mu}\left(\overleftarrow{\partial}_{i}+\overrightarrow{\partial}_{j}\right)_{\mu}+\frac{Pp_{i}}{p_{i}p_{j}}p_{j}^{2}\left(\overrightarrow{\partial}_{j}p_{i}\right)+\frac{Pp_{j}}{p_{i}p_{j}}p_{i}^{2}\left(\overleftarrow{\partial}_{i}p_{j}\right)\right]\, . 
\label{eq:3-2-9}
\eea
\section{Test in two-body elastic scattering}
\label{sec:4}
In this section, we use the result (\ref{eq:3-2-9}) to study a two-body elastic collision. The kinematics for this process is similar to the one shown in  Fig.\ref{FIG:1} with the difference that here we consider all the `hard' particles to be massive and with the same mass $m$.
Thus the energy-momentum conservation condition becomes $E^{2}=p^{2}+k^{2}+m^{2}$ and the Mandelstam variables satisfy the relation: $s+t+u=4m^{2}$. The contribution from the $(i, j)$-pair takes the form 
\bea
\frac{dB_{1}^{(i,j)}}{d\omega}=& \eta_{i}\eta_{j}\frac{G}{\pi}\frac{1-\vec{v}_{i}\cdot\vec{v}_{j}}{1+\vec{v}_{i}\cdot\vec{v}_{j}}\left[\frac{1}{\beta_{ij}}\ln\frac{1+\beta_{ij}}{1-\beta_{ij}}+\frac{2E}{\sqrt{p^{2}+k^{2}}}\ln\frac{E-\sqrt{p^{2}+k^{2}}}{E+\sqrt{p^{2}+k^{2}}}\right] \\
& \times\Bigg\{ E^{2}[\delta_{0}^{\mu}(1+\vec{v}_{i}\cdot\vec{v}_{j})+\delta_{r}^{\mu}(\vec{v}_{i}+\vec{v}_{j})^{r}]\left(\overleftarrow{\partial}_{i}+\overrightarrow{\partial}_{j}\right)_{\mu} \\
& -\frac{m^{2}}{1-\vec{v}_{i}\cdot\vec{v}_{j}}\left[\left(\delta_{0}^{\mu}+\delta_{r}^{\mu}(\vec{v}_{i})^{r}\right)\left(\overrightarrow{\partial}_{j}\right)_{\mu}+\left(\delta_{0}^{\mu}+\delta_{r}^{\mu}(\vec{v}_{j})^{r}\right)\left(\overleftarrow{\partial}_{i}\right)_{\mu}\right]\Bigg\}\, , 
\label{eq:4-1}
\eea
where $\beta_{ij}$ and $|\vec{v}_{i}|$ are given by 
\bes
|\vec{v}_{i}|=\sqrt{\frac{p^{2}+k^{2}}{E^{2}}}\quad \textrm{and}\quad \beta_{ij}\equiv\sqrt{1-\frac{m^{4}}{\left(p_{i}\cdot p_{j}\right)^{2}}}\, . 
\ees

First considering the case $i,j=1,2$ and  $i,j=3,4$, a direct computation gives 
\beas
&\frac{dB_{1}^{(1,2)}}{d\omega}+\frac{dB_{1}^{(2,1)}}{d\omega}\\
=&\frac{2G}{\pi}\left[\frac{E^{2}+p^{2}+k^{2}}{2E\sqrt{p^{2}+k^{2}}}\ln\frac{E^{2}+p^{2}+k^{2}+2E\sqrt{p^{2}+k^{2}}}{E^{2}+p^{2}+k^{2}-2E\sqrt{p^{2}+k^{2}}}+\frac{2E}{\sqrt{p^{2}+k^{2}}}\ln\frac{E-\sqrt{p^{2}+k^{2}}}{E+\sqrt{p^{2}+k^{2}}}\right]\\
& \times \left\{(p^{2}+k^{2})\overleftrightarrow{\partial}_{E}+E\left(p\overleftrightarrow{\partial}_{p}+k\overleftrightarrow{\partial}_{k}\right)\right\}\, ,
\eeas 
and
\beas
&\frac{dB_{1}^{(3,4)}}{d\omega}+\frac{dB_{1}^{(4,3)}}{d\omega}\\
=&-\frac{2G}{\pi}\left[\frac{E^{2}+p^{2}+k^{2}}{2E\sqrt{p^{2}+k^{2}}}\ln\frac{E^{2}+p^{2}+k^{2}+2E\sqrt{p^{2}+k^{2}}}{E^{2}+p^{2}+k^{2}-2E\sqrt{p^{2}+k^{2}}}+\frac{2E}{\sqrt{p^{2}+k^{2}}}\ln\frac{E-\sqrt{p^{2}+k^{2}}}{E+\sqrt{p^{2}+k^{2}}}\right]\\
&\times\Bigg\{(p^{2}+k^{2})\overleftrightarrow{\partial}_{E}+E\left(p\overleftrightarrow{\partial}_{p}+k\overleftrightarrow{\partial}_{k}\right)\Bigg\}\, .
\eeas
Thus, we obtain
\beq
\frac{dB_{1}^{(1,2)}}{d\omega}+\frac{dB_{1}^{(2,1)}}{d\omega}+\frac{dB_{1}^{(3,4)}}{d\omega}+\frac{dB_{1}^{(4,3)}}{d\omega}=0
\label{eq:4-2}
\eeq
which means that the contributions from (1,2) and (3,4) cancel with each other.

For $i,j=1,3$, after a long but straightforward computation, one finds
\beas
\frac{dB_{1}^{(1,3)}}{d\omega}=&-\frac{G}{\pi}\frac{E^{2}+p^{2}-k^{2}}{E^{2}-p^{2}+k^{2}}\left[\frac{E^{2}+p^{2}-k^{2}}{2p\sqrt{E^{2}-k^{2}}}\ln\frac{E^{2}+p^{2}-k^{2}+2p\sqrt{E^{2}-k^{2}}}{E^{2}+p^{2}-k^{2}-2p\sqrt{E^{2}-k^{2}}}\right.\\&\left.+\frac{2E}{\sqrt{p^{2}+k^{2}}}\ln\frac{E-\sqrt{p^{2}+k^{2}}}{E+\sqrt{p^{2}+k^{2}}}\right]\Bigg\{ E^{2}\left[\frac{E^{2}-p^{2}+k^{2}}{E^{2}}\left(\overleftarrow{\partial}_{E}-\overrightarrow{\partial}_{E}\right)+\frac{2k}{E}\left(\overleftarrow{\partial}_{k}-\overrightarrow{\partial}_{k}\right)\right]\\&-\frac{E^{2}m^{2}}{E^{2}+p^{2}-k^{2}}\left[\left(-\overrightarrow{\partial}_{E}+\overleftarrow{\partial}_{E}\right)+\frac{p}{E}\left(\overrightarrow{\partial}_{p}-\overleftarrow{\partial}_{p}\right)+\frac{k}{E}\left(-\overrightarrow{\partial}_{k}+\overleftarrow{\partial}_{k}\right)\right]\Bigg\}\\
=& -\frac{dB_{1}^{(3,1)}}{d\omega} \, .
\eeas
Once again, the total contribution from $(1,3)$-pair vanishes. We have also checked that the same happens for the other pairs: (1,4), (2,3), (2,4), proving that ${dB_{1}}/{d\omega}=0$ is also valid in the case of massive two-body collisions with all particles having the same mass.  
\section{Gravitational elastic scattering}
\label{sec:5}
In this section, we apply the new sub-leading differential $B$-factor obtained in previous sections to the case of a gravitational elastic scattering.

Indeed, from a classical prospective this can correspond to the case of a test light body 
elastically deflected but not captured by a Black Hole of mass $M$.

 For the case in which the test body has a mass $m<<M$, the differential $B$-factor has an asymptotic expression in the limit of $m\rightarrow 0$ (see appendix \ref{appendix:C} for detail derivation)
 as follows 
\beq
\frac{dB_{1}}{d\omega}\sim \frac{GM}{\pi}\left(6+\frac{2}{3}\frac{P^{2}}{M^{2}}\right) \left(\vec{p}_{f}\overleftrightarrow{\partial}_{\vec{p}_{f}}-\vec{p}_{i}\overleftrightarrow{\partial}_{\vec{p}_{i}} \right) \, , 
\label{eq:5-1}
\eeq
where $P=|\vec{p}|$, $M$ is the mass of the black hole, $\vec{p}_{i},\vec{p}_{f}$ are initial and final momenta in the CM frame. Then 
the soft sub-leading radiation in ZFL can be obtained applying the differential B-factor in 
eq. (\ref{eq:5-1}) on the amplitude of gravitational $2\rightarrow 2$ elastic scattering
\beq
\mathcal{M}^{*}\frac{dB_{1}}{d\omega} \mathcal{M}\, ,
\label{eq:5-2}
\eeq
where the partial derivatives with left (right) arrow act on $\mathcal{M}^{*}$ ($\mathcal{M}$).

From field theory prospective, let us 
start from the Einstein-Hilbert action coupled to two scalar fields $\phi_{1}$ and $\phi_{2}$
\beq
\mathcal{S}=\int d^{4}x\sqrt{-g}\left[\frac{1}{16\pi G}R-\frac{1}{2}\partial^{\mu}\phi_{1}\partial_{\mu}\phi_{1}-\frac{1}{2}M^{2}\phi_{1}^{2}-\frac{1}{2}\partial^{\nu}\phi_{2}\partial_{\nu}\phi_{2}-\frac{1}{2}m^{2}\phi_{2}^{2}\right]\, .
\label{eq:5-3}
\eeq

\begin{figure}[H]
\centering
\includegraphics[width=.4\textwidth]{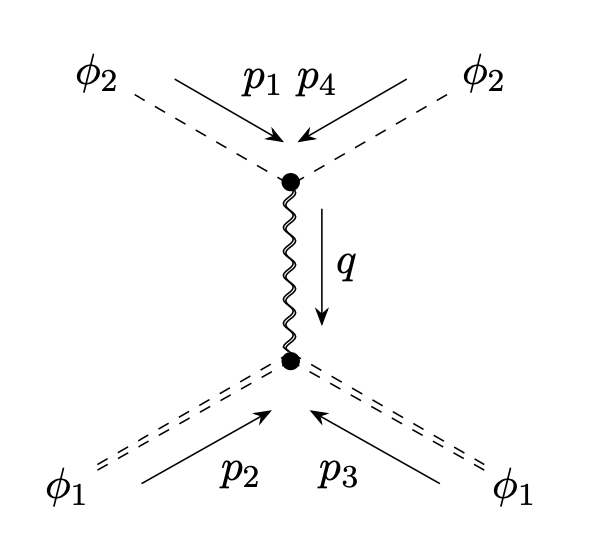}
 \caption{Feynman's diagram of gravitational elastic scattering $\phi_{1}(p_{1})\phi_{2}(p_{2})\rightarrow \phi_{1}(p_{3})\phi_{2}(p_{4})$, 
 where $\phi_{1,2}$ are two scalar fields (dashed lines), the curly line represent the propagator of the graviton field 
 and $p_{1,2,3,4}$ are the four-momenta of in and out states respectively.}
 \label{FIG:2}
\end{figure}

Indeed, for the moment, as an illustrative example, we consider spin-less field gravitational scatterings
(see Fig.\ref{FIG:2}).
The amplitude for this classical tree-level diagram is \cite{7}
\beq
\mathcal{M}=\frac{16 \pi G}{q^{2}}\left(M^{2} m^{2}-2\left(p_{1} \cdot p_{2}\right)^{2}-\left(p_{1} \cdot p_{2}\right) q^{2}\right)\, ,
\label{eq:5-4}
\eeq
where $p_{1},p_{2}$ ($p_{3},p_{4}$) are incoming (outgoing) momenta, $p_{1}^{2}=p_{4}^{2}=-m^{2}$, $p_{2}^{2}=p_{3}^{2}=-M^{2}$ and the momentum transfer is $q=p_{1}+p_{4}=-(p_{2}+p_{3})$. With the kinematics discussed in Appendix \ref{appendix:C}, the amplitude in the center of mass frame becomes 
\begin{align}
\mathcal{M}&= \frac{8\pi G}{(P^{2}+kk^{\prime}-pp^{\prime})}\left[M^{2}m^{2}-2P\left(M+P+\frac{P^{2}}{2M}\right)\left(P\left(M+\frac{P^{2}}{2M}\right)-kk^{\prime}+pp^{\prime}\right)\right] \notag \\
& \sim\frac{8\pi G}{(P^{2}+kk^{\prime}-pp^{\prime})}\left[M^{2}m^{2}-2P\left(M+P\right)\left(MP-kk^{\prime}+pp^{\prime}\right)\right]\, .
\label{eq:5-5}
\end{align}

The derivatives of this amplitude give
\bea
\overrightarrow{\partial}_{p}\mathcal{M}&=\frac{8\pi G}{(P^{2}+kk^{\prime}-pp^{\prime})^{2}}p^{\prime}\left(M^{2}m^{2}-2P^{2}(M+P)^{2}\right)=\mathcal{M}^{*}\overleftarrow{\partial}_{p}\, , \\
\overrightarrow{\partial}_{k}\mathcal{M}&=\frac{8\pi G}{(P^{2}+kk^{\prime}-pp^{\prime})^{2}}k^{\prime}\left(2P^{2}(M+P)^{2}-M^{2}m^{2}\right)=\mathcal{M}^{*}\overleftarrow{\partial}_{k}\, , \\
\overrightarrow{\partial}_{p^{\prime}}\mathcal{M}&=\frac{8\pi G}{(P^{2}+kk^{\prime}-pp^{\prime})^{2}}p\left(M^{2}m^{2}-2P^{2}(M+P)^{2}\right)=\mathcal{M}^{*}\overleftarrow{\partial}_{p^{\prime}}\, , \\
\overrightarrow{\partial}_{k^{\prime}}\mathcal{M}&=\frac{8\pi G}{(P^{2}+kk^{\prime}-pp^{\prime})^{2}}k\left(2P^{2}(M+P)^{2}-M^{2}m^{2}\right)=\mathcal{M}^{*}\overleftarrow{\partial}_{k^{\prime}}\, .
\label{eq:5-6}
\eea

Then it is easy to check that 
\beas
p\left[\left(\mathcal{M}^{*}\overleftarrow{\partial}_{p}\right)\mathcal{M}+\mathcal{M}^{*}\left(\overrightarrow{\partial}_{p}\mathcal{M}\right)\right] - p^{\prime}\left[\left(\mathcal{M}^{*}\overleftarrow{\partial}_{p^{\prime}}\right)\mathcal{M}+\mathcal{M}^{*}\left(\overrightarrow{\partial}_{p^{\prime}}\mathcal{M}\right)\right]=& 0 \, , \\
k\left[\left(\mathcal{M}^{*}\overleftarrow{\partial}_{k}\right)\mathcal{M}+\mathcal{M}^{*}\left(\overrightarrow{\partial}_{k}\mathcal{M}\right)\right] - k^{\prime}\left[\left(\mathcal{M}^{*}\overleftarrow{\partial}_{k^{\prime}}\right)\mathcal{M}+\mathcal{M}^{*}\left(\overrightarrow{\partial}_{k^{\prime}}\mathcal{M}\right)\right]=& 0 \, ,
\eeas
which implies
\beq
\mathcal{M}\left(p^{\prime}\overleftrightarrow{\partial}_{p^{\prime}}+k^{\prime}\overleftrightarrow{\partial}_{k^{\prime}}-p\overleftrightarrow{\partial}_{p}-k\overleftrightarrow{\partial}_{k}\right)\mathcal{M}^{*}=0 \, .
\label{eq:5-7}
\eeq

Thus once again, we get a vanishing result
\beq
\mathcal{M}^{*}\frac{dB_{1}}{d\omega}\mathcal{M}= 0\, ,
\label{eq:5-8}
\eeq
even if the differential $B$-factor is non-zero.
Furthermore, in Appendix \ref{appendix:D}, we find that eq. (\ref{eq:5-8}) is still re-obtained in the case of massive $m$ test particle.
\section{Inelastic scattering}
\label{sec:6}
Here, we will show that ${dB_{1}}/{d\omega}$ gives a finite non-zero result in the $2\rightarrow 2$ inelastic scattering case. 
Indeed, in the case of test particle deflected by a BH, it is reasonable to consider the elastic scattering regime 
only for small deflection angles. In case of harder deflections we would expect that inelastic channels would be opened.

Following similar discussion as Appendix \ref{appendix:C}, but this time take 
\beq
p^{2}+k^{2}=P^{2}\neq P^{\prime 2}=p^{\prime 2}+k^{\prime 2}
\label{eq:6-1}
\eeq
with eq. (\ref{eq:3-2-9}), we can get the contributions from different particle pairs. As shown in Appendix \ref{appendix:E}, however, the result is a long expression.

Nevertheless, in the non-relativistic regime, performing 
a Taylor series expansion up to the second order of $O(P/M)$ and $O({P^{\prime}}/{M^{\prime}})$, the six contributions simplify to
\bes
\frac{dB_{1}^{(1,2)}}{d\omega}+\frac{dB_{1}^{(2,1)}}{d\omega}
\sim \frac{GM}{\pi}\left(-2+\frac{1}{3}\frac{P^{2}}{M^{2}}\right)\left(P\overleftrightarrow{\partial}_{P}+p\overleftrightarrow{\partial}_{p}+k\overleftrightarrow{\partial}_{k}\right)\, , 
\ees
\bes
\frac{dB_{1}^{(3,4)}}{d\omega}+\frac{dB_{1}^{(4,3)}}{d\omega}
\sim \frac{GM^{\prime}}{\pi}\left(2-\frac{1}{3}\frac{P^{\prime2}}{M^{\prime2}}\right)\left(P^{\prime}\overleftrightarrow{\partial}_{P^{\prime}}+p^{\prime}\overleftrightarrow{\partial}_{p^{\prime}}+k^{\prime}\overleftrightarrow{\partial}_{k^{\prime}}\right)\, , 
\ees
\bes
\frac{dB_{1}^{(2,4)}}{d\omega}+\frac{dB_{1}^{(4,2)}}{d\omega}
\sim \frac{GM^{\prime}}{\pi}\left(-2+\frac{1}{3}\frac{P^{\prime2}}{M^{\prime2}}\right)\left(P\overleftrightarrow{\partial}_{P}+p\overleftrightarrow{\partial}_{p}+k\overleftrightarrow{\partial}_{k}\right)\, , 
\ees
\bes
\frac{dB_{1}^{(1,3)}}{d\omega}+\frac{dB_{1}^{(3,1)}}{d\omega}
\sim \frac{GM}{\pi}\left(-2+\frac{1}{3}\frac{P^{2}}{M^{2}}\right)\left(-P^{\prime}\overleftrightarrow{\partial}_{P^{\prime}}-p^{\prime}\overleftrightarrow{\partial}_{p^{\prime}}-k^{\prime}\overleftrightarrow{\partial}_{k^{\prime}}\right)\, , 
\ees
\bes
\frac{dB_{1}^{(1,4)}}{d\omega}+\frac{dB_{1}^{(4,1)}}{d\omega}
\sim \frac{G}{\pi}\Bigg[-2-\frac{2}{3}\left(\frac{P^{2}}{M^{2}}+\frac{P^{\prime2}}{M^{\prime2}}\right)\Bigg]\left(pM^{\prime}\overleftrightarrow{\partial}_{p}+kM^{\prime}\overleftrightarrow{\partial}_{k}-p^{\prime}M\overleftrightarrow{\partial}_{p^{\prime}}-k^{\prime}M\overleftrightarrow{\partial}_{k^{\prime}}\right) \, , 
\ees
\beas
\frac{dB_{1}^{(2,3)}}{d\omega}+\frac{dB_{1}^{(3,2)}}{d\omega}=&\frac{2G}{\pi}\frac{PP^{\prime}-(pp^{\prime}-kk^{\prime})}{PP^{\prime}+(pp^{\prime}-kk^{\prime})}\ln\left[\frac{PP^{\prime}-(pp^{\prime}-kk^{\prime})}{2PP^{\prime}}\right]\Bigg\{\left[PP^{\prime}+(pp^{\prime}-kk^{\prime})\right]\\&\times\left(\overleftrightarrow{\partial}_{P}-\overleftrightarrow{\partial}_{P^{\prime}}\right)+\left(pP^{\prime}+p^{\prime}P\right)\left(\overleftrightarrow{\partial}_{p}-\overleftrightarrow{\partial}_{p^{\prime}}\right)+\left(kP^{\prime}-k^{\prime}P\right)\left(\overleftrightarrow{\partial}_{k}+\overleftrightarrow{\partial}_{k^{\prime}}\right)\Bigg\}\, .
\eeas

Summing the single contributions, we find that the total sub-leading $B$-fractor for the inelastic scattering is dominated by the following leading term:
\bea
\frac{dB_{1}}{d\omega}=&\left[\frac{GM}{\pi}\left(2-\frac{1}{3}\frac{P^{2}}{M^{2}}\right)+\frac{GM^{\prime}}{\pi}\left(2-\frac{1}{3}\frac{P^{\prime2}}{M^{\prime2}}\right)\right]\left(P^{\prime}\overleftrightarrow{\partial}_{P^{\prime}}-P\overleftrightarrow{\partial}_{P}+p^{\prime}\overleftrightarrow{\partial}_{p^{\prime}}\right.\\&\left.-p\overleftrightarrow{\partial}_{p}+k^{\prime}\overleftrightarrow{\partial}_{k^{\prime}}-k\overleftrightarrow{\partial}_{k}\right)+\frac{G}{\pi}\Bigg[-2-\frac{2}{3}\left(\frac{P^{2}}{M^{2}}+\frac{P^{\prime2}}{M^{\prime2}}\right)\Bigg]\\&\times\left[M^{\prime}\left(p\overleftrightarrow{\partial}_{p}+k\overleftrightarrow{\partial}_{k}\right)-M\left(p^{\prime}\overleftrightarrow{\partial}_{p^{\prime}}-k^{\prime}\overleftrightarrow{\partial}_{k^{\prime}}\right)\right]\, .
\label{eq:6-2}
\eea
\section{Conclusions and remarks}
\label{sec:conclusions}
In this paper, we computed the sub-leading order expansion of the
gravitational radiation energy spectrum $dE^{GW}/d\omega$
in the ZFL for
a generic multi-body or multi-particle collision. 
Our result can be applied on both massless and massive fundamental/composite particle scatterings
with every spin as well as to bodies such as Black Holes. 

As an application, we considered the case of 
sub-leading soft emission from a
 gravitational elastic scattering.
 From the point of view of BH physics, it can be the case of 
 a test body deflected, but not captured, by a BH. 
 
 Surprisingly, we obtain that all sub-leading terms of 
 the energy radiation spectrum, emitted from an elastic collision, 
 exactly cancel each others, in both massless and massive particle collisions. 
 On the other hand, for inelastic $2\rightarrow 2$ scatterings, we obtain 
 a general and complicated non-zero analytic result 
which may be applied in the case of GW soft emission from BH gravitational inelastic scatterings.  
It may certainly be an attractive possibility to relate our result on 
 searches of BMS gravitational memory effects from GW physics. 
In this sense, our results may be applied in BH mergings  seen as a gravitational BH capture inelastic scattering process. 
On the other hand, violations of universality of our results from loop radiative corrections would 
be expected from previous analysis as mentioned in the introduction above. 

Another intriguing possibility is to consider {\it footprints} of string theory 
effects on leading and sub-leading gravitational radiation from $\alpha'$ corrections and Regge poles that may survive with a polynomial decay in radius,
dubbed string memory effect \cite{Addazi:2020obs}.

Finally, the very next step for ZFL program would be to compute 
sub-sub-leading order on the $dE^{GW}/d\omega$ expansion in $\omega$. 
In Ref.\cite{ABV} the case of massless particle high energy collisions 
was considered 
-- having in mind an application to transplanckian scattering regimes.
However a more general result is missing yet and it is beyond the purpose of this paper. 

\section*{Acknowledgments}
AA would like to thank Massimo Bianchi for interesting and useful discussions and remarks on this subject. 
Our work is supported by the Talent Scientific Research Program of College of Physics, Sichuan University, Grant No.1082204112427
\& the Fostering Program in Disciplines Possessing Novel Features for Natural Science of Sichuan University,  Grant No. 2020SCUNL209
\& 1000 Talent program of Sichuan province 2021. 

\appendix

\section{Gravitational elastic scattering I}
\label{appendix:C}
In this appendix, we give a detail computation of the differential $B$-factor for the gravitational elastic scattering process with the test particle being massless as discussed in section \ref{sec:4}.
 In the center of mass frame, the kinematical variables of this process are
\bea
p_{1}&= \left(M+\frac{P^{2}}{2M},p,k,0\right) \, ,\\
p_{2}&= (P,-p,-k,0) \, ,\\
p_{3}&= (-P,p^{\prime},-k^{\prime},0) \, ,\\
p_{4}&= \left(-M-\frac{P^{2}}{2M},-p^{\prime},k^{\prime},0\right)\, ,
\label{eq:C-1}
\eea
where $p,p^{\prime},k,k^{\prime}$ satisfy the realtion
\bes
p^{2}+k^{2}=P^{2}=p^{\prime2}+k^{\prime2}\, .
\ees

Some properties relating eq.(\ref{eq:3-2-9}) are the following
\beas 
P=& p_{1}+p_{2}=\left(M+P+\frac{P^{2}}{2M},\vec{0}\right) \, , \\
s=& -(p_{1}+p_{2})^{2}=\left(M+P+\frac{P^{2}}{2M}\right)^{2} \, , \\
\tilde{s}_{ij}=& -\left(M+P+\frac{P^{2}}{2M}\right)^{2}\frac{1+\vec{v}_{i}\cdot\vec{v}_{j}}{1-\vec{v}_{i}\cdot\vec{v}_{j}} \, , \\
Q_{ij}^{\mu}=& -\left(M+P+\frac{P^{2}}{2M}\right)\left(\frac{1+\vec{v}_{i}\cdot\vec{v}_{j}}{1-\vec{v}_{i}\cdot\vec{v}_{j}},\frac{\vec{v}_{i}+\vec{v}_{j}}{1-\vec{v}_{i}\cdot\vec{v}_{j}}\right)\, ,
\eeas
and the velocities are given by 
\begin{alignat*}{5} 
\vec{v}_{1}&=\frac{1}{M}(p,k,0)\, , &\qquad \vec{v}_{2}=&\frac{1}{P}(-p,-k,0) \, ,&\qquad |\vec{v}_{1}|=& \frac{P}{M} \, ,&\qquad |\vec{v}_{2}|=& \frac{P}{\sqrt{m^{2}+P^{2}}}\rightarrow1\, , \\
\vec{v}_{3}&=\frac{1}{P}(-p^{\prime},k^{\prime},0) \, ,&\qquad \vec{v}_{4}=&\frac{1}{M}(p^{\prime},-k^{\prime},0) \, ,&\qquad |\vec{v}_{3}|=& \frac{P}{\sqrt{m^{2}+P^{2}}}\rightarrow1 \, ,&\qquad |\vec{v}_{4}|=&\frac{P}{M} \, .\\
\end{alignat*} 

Their are six $(i,j)$-pairs we need to take into consideration, namely: $(1,2)$, $(1,3)$, $(1,4)$, $(2,3)$, $(2,4)$, $(3,4)$. After some tedious but straightforward computations, we obtain the contributions from $(1,2)+(3,4)$:
\beas
&\frac{dB_{1}^{(1,2)}}{d\omega}+\frac{dB_{1}^{(2,1)}}{d\omega}+\frac{dB_{1}^{(3,4)}}{d\omega}+\frac{dB_{1}^{(4,3)}}{d\omega}\\=&\frac{G}{\pi}\left[\ln\left(\frac{M+P+\frac{P^{2}}{2M}}{M}\right)^{2}+\frac{M}{P}\ln\frac{M-P}{M+P}\right]\frac{M+P}{M-P}\left(P-M-\frac{P^{2}}{2M}\right)\\
& \times \left(p^{\prime}\overleftrightarrow{\partial}_{p^{\prime}}+k^{\prime}\overleftrightarrow{\partial}_{k^{\prime}}-p\overleftrightarrow{\partial}_{p}-k\overleftrightarrow{\partial}_{k}\right)\, , 
\eeas
and from $(1,3)+(2,4)$:
\beas
&\frac{dB_{1}^{(1,3)}}{d\omega}+\frac{dB_{1}^{(3,1)}}{d\omega}+\frac{dB_{1}^{(2,4)}}{d\omega}+\frac{dB_{1}^{(4,2)}}{d\omega}\\=&\frac{G}{\pi}\frac{PM+(pp^{\prime}-kk^{\prime})}{PM-(pp^{\prime}-kk^{\prime})}\left[\ln\left(\frac{P\left(M+\frac{P^{2}}{2M}\right)+pp^{\prime}-kk^{\prime}}{MP}\right)^{2}+\frac{M}{P}\ln\frac{M-P}{M+P}\right]\\&\times\Bigg\{\left(P\left(M+\frac{P^{2}}{2M}\right)+(pp^{\prime}-kk^{\prime})\right)\Bigg[\frac{(p-p^{\prime})(P+M)}{PM+(pp^{\prime}-kk^{\prime})}\left(\overleftrightarrow{\partial}_{p}+\overleftrightarrow{\partial}_{p^{\prime}}\right)\\&+\frac{(k+k^{\prime})(P+M)}{PM+(pp^{\prime}-kk^{\prime})}\left(\overleftrightarrow{\partial}_{k}-\overleftrightarrow{\partial}_{k^{\prime}}\right)\Bigg]+\frac{P\left[P^{2}-\left(M+\frac{P^{2}}{2M}\right)^{2}\right]}{P\left(M+\frac{P^{2}}{2M}\right)+pp^{\prime}-kk^{\prime}}\\&\times\left(p\overleftrightarrow{\partial}_{p^{\prime}}-k\overleftrightarrow{\partial}_{k^{\prime}}-p^{\prime}\overleftrightarrow{\partial}_{p}+k^{\prime}\overleftrightarrow{\partial}_{k}\right)\Bigg\}\, , 
\eeas
from $(1,4)$:
\beas
&\frac{dB_{1}^{(1,4)}}{d\omega}+\frac{dB_{1}^{(4,1)}}{d\omega}\\
=&\frac{G}{\pi}\Bigg[\frac{M}{\sqrt{2}\sqrt{P^{2}-(pp^{\prime}-kk^{\prime})}}\ln\frac{1+\sqrt{2}\frac{\sqrt{P^{2}-(pp^{\prime}-kk^{\prime})}}{M}}{1-\sqrt{2}\frac{\sqrt{P^{2}-(pp^{\prime}-kk^{\prime})}}{M}}+2\frac{M}{P}\ln\frac{M-P}{M+P}\Bigg]M
\\&\times\left(p\overleftrightarrow{\partial}_{p}-p^{\prime}\overleftrightarrow{\partial}_{p^{\prime}}+k\overleftrightarrow{\partial}_{k}-k^{\prime}\overleftrightarrow{\partial}_{k^{\prime}}\right)\, , 
\eeas
and finally from $(2,3)$:
\beas
&\frac{dB_{1}^{(2,3)}}{d\omega}+\frac{dB_{1}^{(3,2)}}{d\omega}\\=&\frac{G}{\pi}\frac{P^{2}-(pp^{\prime}-kk^{\prime})}{P^{2}+(pp^{\prime}-kk^{\prime})}\ln\left[\frac{P^{2}-(pp^{\prime}-kk^{\prime})}{2P^{2}}\right]^{2}\left[P(p+p^{\prime})\left(\overleftrightarrow{\partial}_{p}-\overleftrightarrow{\partial}_{p^{\prime}}\right)-P(k^{\prime}-k)\left(\overleftrightarrow{\partial}_{k}+\overleftrightarrow{\partial}_{k^{\prime}}\right)\right]\, .
\eeas

To get a compact and meaningful differential $B$-factor, we do the Taylor series expansion up to the second order at $P/M$. Then the above results are simplified to 
\bes
\frac{dB_{1}^{(1,2)}}{d\omega}+\frac{dB_{1}^{(2,1)}}{d\omega}+\frac{dB_{1}^{(3,4)}}{d\omega}+\frac{dB_{1}^{(4,3)}}{d\omega}
\sim \frac{GM}{\pi}\left(2-\frac{1}{3}\frac{P^{2}}{M^{2}}\right)\left(p^{\prime}\overleftrightarrow{\partial}_{p^{\prime}}+k^{\prime}\overleftrightarrow{\partial}_{k^{\prime}}-p\overleftrightarrow{\partial}_{p}-k\overleftrightarrow{\partial}_{k}\right)\, , 
\ees
\bes
\frac{dB_{1}^{(1,3)}}{d\omega}+\frac{dB_{1}^{(3,1)}}{d\omega}+\frac{dB_{1}^{(2,4)}}{d\omega}+\frac{dB_{1}^{(4,2)}}{d\omega} \sim \frac{GM}{\pi}\left(-2+\frac{1}{3}\frac{P^{2}}{M^{2}}\right)\left(p\overleftrightarrow{\partial}_{p}-p^{\prime}\overleftrightarrow{\partial}_{p^{\prime}}+k\overleftrightarrow{\partial}_{k}-k^{\prime}\overleftrightarrow{\partial}_{k^{\prime}}\right)\, , 
\ees
and
\beas
\frac{dB_{1}^{(1,4)}}{d\omega}+\frac{dB_{1}^{(4,1)}}{d\omega}=&\frac{2G}{\pi}\Bigg[1+\frac{M}{P}\ln\frac{1-\frac{P}{M}}{1+\frac{P}{M}}\Bigg]M\left(p\overleftrightarrow{\partial}_{p}-p^{\prime}\overleftrightarrow{\partial}_{p^{\prime}}+k\overleftrightarrow{\partial}_{k}-k^{\prime}\overleftrightarrow{\partial}_{k^{\prime}}\right)\\\sim&\frac{2G}{\pi}\Bigg[-1-\frac{2}{3}\frac{P^{2}}{M^{2}}\Bigg]M\left(p\overleftrightarrow{\partial}_{p}-p^{\prime}\overleftrightarrow{\partial}_{p^{\prime}}+k\overleftrightarrow{\partial}_{k}-k^{\prime}\overleftrightarrow{\partial}_{k^{\prime}}\right)\, . 
\eeas

Summing up the above results, we find that the total differential $B$-factor: ${dB_{1}}/{d\omega}=\sum^{i\neq j}_{i,j} {dB_{1}^{(i,j)}}/{d\omega}$ has a leading expression as 
\beas
\frac{dB_{1}}{d\omega}\sim&\frac{GM}{\pi}\left(6+\frac{2}{3}\frac{P^{2}}{M^{2}}\right)\left(p^{\prime}\overleftrightarrow{\partial}_{p^{\prime}}+k^{\prime}\overleftrightarrow{\partial}_{k^{\prime}}-p\overleftrightarrow{\partial}_{p}-k\overleftrightarrow{\partial}_{k}\right)\\=&\frac{GM}{\pi}\left(6+\frac{2}{3}\frac{P^{2}}{M^{2}}\right)\left(\vec{p}_{f}\overleftrightarrow{\partial}_{\vec{p}_{f}}-\vec{p}_{i}\overleftrightarrow{\partial}_{\vec{p}_{i}}\right) \, , 
\label{eq:C-2}
\eeas
where $\vec{p}_{i}$ ($\vec{p}_{f}$) are momenta of initial (final) states and $\overleftrightarrow{\partial}\equiv \overleftarrow{\partial}+\overrightarrow{\partial}$.
\section{Gravitational elastic scattering II}
\label{appendix:D}
Our goal in this appendix is to generalize the discussion in Appendix \ref{appendix:C} to the case in which the test particle has mass $m$.
 The kinematics is similar to the previous case and reads
\bea
p_{1}&=\left(M+\frac{P^{2}}{2M},p,k,0\right)\, ,\\
p_{2}&=(\sqrt{P^{2}+m^{2}},-p,-k,0)\, , \\
p_{3}&=(-\sqrt{P^{2}+m^{2}},p^{\prime},-k^{\prime},0)\, ,  \\
p_{4}&=\left(-M-\frac{P^{2}}{2M},-p^{\prime},k^{\prime},0\right)\, , 
\label{eq:D-1}
\eea
with $p^{2}+k^{2}=P^{2}=p^{\prime 2}+k^{\prime 2}$. 

Again, let us consider the contributions from six pairs separately: first for $(1,2)+(3,4)$, considering that $\tilde{s}_{12}=\tilde{s}_{34}$ and $p_{1}\cdot p_{2}=p_{3}\cdot p_{4}$, 
we obtain 
\beas
\frac{dB_{1}^{(1,2)}}{d\omega}+\frac{dB_{1}^{(2,1)}}{d\omega}=&-\frac{G\sqrt{s}}{\pi}\frac{1}{\tilde{s}_{12}}\left[\frac{1}{\beta_{12}}\ln\frac{1+\beta_{12}}{1-\beta_{12}}+\left(\frac{1}{\left|\vec{v}_{1}\right|}\ln\frac{1-\left|\vec{v}_{1}\right|}{1+\left|\vec{v}_{1}\right|}+\frac{1}{\left|\vec{v}_{2}\right|}\ln\frac{1-\left|\vec{v}_{2}\right|}{1+\left|\vec{v}_{2}\right|}\right)\right]\\&\times\left[(p_{1}p_{2})Q_{12}^{\mu}\left(\overleftrightarrow{\partial}_{1}+\overleftrightarrow{\partial}_{2}\right)_{\mu}+\frac{Pp_{1}}{p_{1}p_{2}}p_{2}^{2}\overleftrightarrow{\partial}_{2}p_{1}+\frac{Pp_{2}}{p_{1}p_{2}}p_{1}^{2}\overleftrightarrow{\partial}_{1}p_{2}\right] \, , 
\eeas
and
\beas
\frac{dB_{1}^{(3,4)}}{d\omega}+\frac{dB_{1}^{(4,3)}}{d\omega}=&-\frac{G\sqrt{s}}{\pi}\frac{1}{\tilde{s}_{34}}\left[\frac{1}{\beta_{34}}\ln\frac{1+\beta_{34}}{1-\beta_{34}}+\left(\frac{1}{\left|\vec{v}_{3}\right|}\ln\frac{1-\left|\vec{v}_{3}\right|}{1+\left|\vec{v}_{3}\right|}+\frac{1}{\left|\vec{v}_{4}\right|}\ln\frac{1-\left|\vec{v}_{4}\right|}{1+\left|\vec{v}_{4}\right|}\right)\right]\\&\times\left[(p_{3}p_{4})Q_{34}^{\mu}\left(\overleftrightarrow{\partial}_{3}+\overleftrightarrow{\partial}_{4}\right)_{\mu}+\frac{Pp_{3}}{p_{3}p_{4}}p_{4}^{2}\overleftrightarrow{\partial}_{4}p_{3}+\frac{Pp_{4}}{p_{3}p_{4}}p_{3}^{2}\overleftrightarrow{\partial}_{3}p_{4}\right] \, , 
\eeas
where the $Q^{\mu}$ vectors are given by
\beas
Q_{12}^{\mu}&=-\frac{M+\frac{P^{2}}{2M}+\sqrt{P^{2}+m^{2}}}{M\sqrt{P^{2}+m^{2}}+P^{2}}\left(M\sqrt{P^{2}+m^{2}}-P^{2},(\sqrt{P^{2}+m^{2}}-M)\left(p,k,0\right)\right)\, , \\
Q_{34}^{\mu}&=-\frac{M+\frac{P^{2}}{2M}+\sqrt{P^{2}+m^{2}}}{M\sqrt{P^{2}+m^{2}}+P^{2}}\left(M\sqrt{P^{2}+m^{2}}-P^{2},(\sqrt{P^{2}+m^{2}}-M)\left(p^{\prime},-k^{\prime},0\right)\right) \, .
\eeas

Thus we have
\beas
&\frac{dB_{1}^{(1,2)}}{d\omega}+\frac{dB_{1}^{(2,1)}}{d\omega}+\frac{dB_{1}^{(3,4)}}{d\omega}+\frac{dB_{1}^{(4,3)}}{d\omega}\\=&-\frac{G\sqrt{s}}{\pi}\frac{1}{\tilde{s}_{12}}\left[\frac{1}{\beta_{12}}\ln\frac{1+\beta_{12}}{1-\beta_{12}}+\left(\frac{1}{\left|\vec{v}_{1}\right|}\ln\frac{1-\left|\vec{v}_{1}\right|}{1+\left|\vec{v}_{1}\right|}+\frac{1}{\left|\vec{v}_{2}\right|}\ln\frac{1-\left|\vec{v}_{2}\right|}{1+\left|\vec{v}_{2}\right|}\right)\right]\\&\times\Bigg\{\frac{Pp_{1}}{p_{1}p_{2}}m^{2}\left(p\overleftrightarrow{\partial}_{p}+k\overleftrightarrow{\partial}_{k}-p^{\prime}\overleftrightarrow{\partial}_{p^{\prime}}-k^{\prime}\overleftrightarrow{\partial}_{k^{\prime}}\right)+\frac{Pp_{2}}{p_{1}p_{2}}\left[P^{2}-\left(M+\frac{P^{2}}{2M}\right)^{2}\right]\\&\times\left(p^{\prime}\overleftrightarrow{\partial}_{p^{\prime}}+k^{\prime}\overleftrightarrow{\partial}_{k^{\prime}}-p\overleftrightarrow{\partial}_{p}-k\overleftrightarrow{\partial}_{k}\right)\Bigg\}\, .
\eeas

The combination: $p^{\prime}\overleftrightarrow{\partial}_{p^{\prime}}+k^{\prime}\overleftrightarrow{\partial}_{k^{\prime}}-p\overleftrightarrow{\partial}_{p}-k\overleftrightarrow{\partial}_{k}$ factorized implies that $\mathcal{M}^{*}\left[(1,2)+(3,4)\right]\mathcal{M}=0$ (see eq. (\ref{eq:5-7})).

Then for $(1,3)+(2,4)$, similarly, considering that $\tilde{s}_{13}=\tilde{s}_{24}, p_{1}\cdot p_{3}=p_{2}\cdot p_{4}$, 
we obtain 
\beas
&\frac{dB_{1}^{(1,3)}}{d\omega}+\frac{dB_{1}^{(3,1)}}{d\omega}+\frac{dB_{1}^{(2,4)}}{d\omega}+\frac{dB_{1}^{(4,2)}}{d\omega}\\=&\frac{G\sqrt{s}}{\pi}\frac{1}{\tilde{s}_{13}}\left[\frac{1}{\beta_{13}}\ln\frac{1+\beta_{13}}{1-\beta_{13}}+\left(\frac{1}{\left|\vec{v}_{1}\right|}\ln\frac{1-\left|\vec{v}_{1}\right|}{1+\left|\vec{v}_{1}\right|}+\frac{1}{\left|\vec{v}_{3}\right|}\ln\frac{1-\left|\vec{v}_{3}\right|}{1+\left|\vec{v}_{3}\right|}\right)\right]\\&\times\Bigg\{-(p_{1}p_{3})\frac{\left(M+\frac{P^{2}}{2M}+\sqrt{P^{2}+m^{2}}\right)\left(\sqrt{P^{2}+m^{2}}+M\right)}{M\sqrt{P^{2}+m^{2}}-(kk^{\prime}-pp^{\prime})}\Bigg(p\overleftrightarrow{\partial}_{p}-p^{\prime}\overleftrightarrow{\partial}_{p^{\prime}}+k\overleftrightarrow{\partial}_{k}-k^{\prime}\overleftrightarrow{\partial}_{k^{\prime}}\\&-p^{\prime}\overleftrightarrow{\partial}_{p}+p\overleftrightarrow{\partial}_{p^{\prime}}+k^{\prime}\overleftrightarrow{\partial}_{k}-k\overleftrightarrow{\partial}_{k^{\prime}}\Bigg)+\frac{Pp_{1}}{p_{1}p_{3}}m^{2}\left[p^{\prime}\overleftrightarrow{\partial}_{p}-p\overleftrightarrow{\partial}_{p^{\prime}}+k\overleftrightarrow{\partial}_{k^{\prime}}-k^{\prime}\overleftrightarrow{\partial}_{k}\right]\\&+\frac{Pp_{2}}{p_{1}p_{3}}\left[P^{2}-\left(M+\frac{P^{2}}{2M}\right)^{2}\right]\left(p\overleftrightarrow{\partial}_{p^{\prime}}-p^{\prime}\overleftrightarrow{\partial}_{p}+k^{\prime}\overleftrightarrow{\partial}_{k}-k\overleftrightarrow{\partial}_{k^{\prime}}\right)\Bigg\}\, . 
\eeas

Let us now note that all the three terms in the above expression contain the combination $p\overleftrightarrow{\partial}_{p^{\prime}}-p^{\prime}\overleftrightarrow{\partial}_{p}+k^{\prime}\overleftrightarrow{\partial}_{k}-k\overleftrightarrow{\partial}_{k^{\prime}}$, which guarantees the cancellation after the application of the expression on the amplitudes
\beas
&\mathcal{M}^{*}\left[-p^{\prime}\overleftrightarrow{\partial}_{p}+p\overleftrightarrow{\partial}_{p^{\prime}}+k^{\prime}\overleftrightarrow{\partial}_{k}-k\overleftrightarrow{\partial}_{k^{\prime}}\right]\mathcal{M}\\
=&\frac{16\pi G\left(2P^{2}(M+P)^{2}-M^{2}m^{2}\right)}{\left(P^{2}+kk^{\prime}-pp^{\prime}\right)^{2}}\left(p^{\prime2}-p^{2}+k^{\prime2}-k^{2}\right)\mathcal{M}=0\, ,
\eeas
thus $\mathcal{M}^{*}\left[(1,3)+(2,4)\right]\mathcal{M}=0$.

Similarly, the final expression for the contribution from $(2,3)$-pair is
\beas
&\frac{dB_{1}^{(2,3)}}{d\omega}+\frac{dB_{1}^{(3,2)}}{d\omega}\\
=&\frac{G\sqrt{s}}{\pi}\frac{1}{\tilde{s}_{23}}\left[\frac{1}{\beta_{23}}\ln\frac{1+\beta_{23}}{1-\beta_{23}}+\frac{2}{\left|\vec{v}_{2}\right|}\ln\frac{1-\left|\vec{v}_{2}\right|}{1+\left|\vec{v}_{2}\right|}\right]\Bigg\{-\left(M+\frac{P^{2}}{2M}+\sqrt{P^{2}+m^{2}}\right)\\&\times\sqrt{P^{2}+m^{2}}\left(p\overleftrightarrow{\partial}_{p}-p^{\prime}\overleftrightarrow{\partial}_{p^{\prime}}+k\overleftrightarrow{\partial}_{k}-k^{\prime}\overleftrightarrow{\partial}_{k^{\prime}}+p^{\prime}\overleftrightarrow{\partial}_{p}-p\overleftrightarrow{\partial}_{p^{\prime}}-k^{\prime}\overleftrightarrow{\partial}_{k}+k\overleftrightarrow{\partial}_{k^{\prime}}\right)\\&-\frac{\left(\sqrt{P^{2}+m^{2}}\right)\left(M+\frac{P^{2}}{2M}+\sqrt{P^{2}+m^{2}}\right)}{(P^{2}+m^{2})-(pp^{\prime}-kk^{\prime})}m^{2}\left(p\overleftrightarrow{\partial}_{p^{\prime}}-p^{\prime}\overleftrightarrow{\partial}_{p}+k^{\prime}\overleftrightarrow{\partial}_{k}-k\overleftrightarrow{\partial}_{k^{\prime}}\right)\Bigg\}
\eeas 
and from $(1,4)$-pair
\beas
&\frac{dB_{1}^{(1,4)}}{d\omega}+\frac{dB_{1}^{(4,1)}}{d\omega}\\=&\frac{G}{\pi}\Bigg[\frac{\left(M+\frac{P^{2}}{2M}\right)^{2}-(pp^{\prime}-kk^{\prime})}{\sqrt{(2M^{2}+P^{2}+\frac{P^{4}}{4M^{2}})-(pp^{\prime}-kk^{\prime})}\sqrt{(P^{2}+\frac{P^{4}}{4M^{2}})-(pp^{\prime}-kk^{\prime})}}\\&\times\ln\frac{\left(M+\frac{P^{2}}{2M}\right)^{2}-(pp^{\prime}-kk^{\prime})+\sqrt{(2M^{2}+P^{2}+\frac{P^{4}}{4M^{2}})-(pp^{\prime}-kk^{\prime})}\sqrt{(P^{2}+\frac{P^{4}}{4M^{2}})-(pp^{\prime}-kk^{\prime})}}{\left(M+\frac{P^{2}}{2M}\right)^{2}-(pp^{\prime}-kk^{\prime})-\sqrt{(2M^{2}+P^{2}+\frac{P^{4}}{4M^{2}})-(pp^{\prime}-kk^{\prime})}\sqrt{(P^{2}+\frac{P^{4}}{4M^{2}})-(pp^{\prime}-kk^{\prime})}}\\&+2\frac{M}{P}\ln\frac{M-P}{M+P}\Bigg]\Bigg\{\left[\left(M+\frac{P^{2}}{2M}\right)^{2}-(pp^{\prime}-kk^{\prime})\right]\left(\frac{M}{M^{2}-(pp^{\prime}-kk^{\prime})}\right)\\&\times\left(p\overleftrightarrow{\partial}_{p}+k\overleftrightarrow{\partial}_{k}-p^{\prime}\overleftrightarrow{\partial}_{p^{\prime}}-k^{\prime}\overleftrightarrow{\partial}_{k^{\prime}}+k\overleftrightarrow{\partial}_{k^{\prime}}-p\overleftrightarrow{\partial}_{p^{\prime}}+p^{\prime}\overleftrightarrow{\partial}_{p}-k^{\prime}\overleftrightarrow{\partial}_{k}\right)\\&+\frac{\left(M+\frac{P^{2}}{2M}\right)\left[P^{2}-\left(M+\frac{P^{2}}{2M}\right)^{2}\right]}{\left(M+\frac{P^{2}}{2M}\right)^{2}-(pp^{\prime}-kk^{\prime})}\left(k\overleftrightarrow{\partial}_{k^{\prime}}-p\overleftrightarrow{\partial}_{p^{\prime}}+p^{\prime}\overleftrightarrow{\partial}_{p}-k^{\prime}\overleftrightarrow{\partial}_{k}\right)\Bigg]\Bigg\}\, .
\eeas
Once again, these terms will give us a zero result after applying them to the amplitude.  

In conclusion, we have generalized $\mathcal{M}^{*}\frac{dB_{1}}{d\omega} \mathcal{M}=0$ to all gravitational elastic scatterings.
\section{$\frac{dB_{1}^{(i,j)}}{d \omega}$ for the inelastic case}
\label{appendix:E}
In this appendix we show the general results (before expansion) of the ${dB_{1}^{(i,j)}}/{d\omega}$ factor for the inelastic scattering discussed in section \ref{sec:6}. Adopting the kinematics (\ref{eq:6-1}), one gets from eq. (\ref{eq:3-2-9}) contributions from six pairs respectively and the results read\footnote{The computation for the inelastic case is similar to those in appendix \ref{appendix:C} and \ref{appendix:D} (although much more complicated), thus to avoid repeatability we only show the results in this appendix.} 
\beas
\frac{dB_{1}^{(1,2)}}{d\omega}+\frac{dB_{1}^{(2,1)}}{d\omega}=&\frac{G}{\pi}\left[\ln\left(\frac{M+\frac{P^{2}}{2M}+P}{M}\right)^{2}+\frac{M}{P}\ln\frac{M-P}{M+P}\right]\Bigg[\left(M+\frac{P^{2}}{2M}+P\right)\\&\times\left(P\overleftrightarrow{\partial}_{M+\frac{P^{2}}{2M}}+P\overleftrightarrow{\partial}_{P}\right)+\left(P-M-\frac{P^{2}}{2M}\right)\frac{M+P}{M-P}\\&\times\left(P\overleftrightarrow{\partial}_{M+\frac{P^{2}}{2M}}-p\overleftrightarrow{\partial}_{p}-k\overleftrightarrow{\partial}_{k}\right)\Bigg]\, , 
\eeas
\beas
\frac{dB_{1}^{(3,4)}}{d\omega}+\frac{dB_{1}^{(4,3)}}{d\omega}=&\frac{G}{\pi}\left[\ln\left(\frac{M^{\prime}+P^{\prime}+\frac{P^{\prime2}}{2M}}{M^{\prime}}\right)^{2}+\frac{M^{\prime}}{P^{\prime}}\ln\frac{M^{\prime}-P^{\prime}}{M^{\prime}+P^{\prime}}\right]\Bigg[-\left(M^{\prime}+P^{\prime}+\frac{P^{\prime2}}{2M^{\prime}}\right)\\&\times\left(P^{\prime}\overleftrightarrow{\partial}_{P^{\prime}}+P^{\prime}\overleftrightarrow{\partial}_{M^{\prime}+\frac{P^{\prime2}}{2M^{\prime}}}\right)-\left(P^{\prime}-M^{\prime}-\frac{P^{\prime2}}{2M^{\prime}}\right)\frac{M^{\prime}+P^{\prime}}{M^{\prime}-P^{\prime}}\\&\times\left(P^{\prime}\overleftrightarrow{\partial}_{M^{\prime}+\frac{P^{\prime2}}{2M^{\prime}}}-p^{\prime}\overleftrightarrow{\partial}_{p^{\prime}}-k^{\prime}\overleftrightarrow{\partial}_{k^{\prime}}\right)\Bigg]\, , 
\eeas
\beas
\frac{dB_{1}^{(1,3)}}{d\omega}+\frac{dB_{1}^{(3,1)}}{d\omega}=&\frac{G}{\pi}\frac{P^{\prime}M+(pp^{\prime}-kk^{\prime})}{P^{\prime}M-(pp^{\prime}-kk^{\prime})}\Bigg[\ln\left(\frac{P^{\prime}\left(M+\frac{P^{2}}{2M}\right)+pp^{\prime}-kk^{\prime}}{MP^{\prime}}\right)^{2}\\&+\frac{M}{P}\ln\frac{M-P}{M+P}\Bigg]\Bigg\{\Bigg[P^{\prime}\left(M+\frac{P^{2}}{2M}\right)+(pp^{\prime}-kk^{\prime})\Bigg(\frac{P^{\prime}M-(pp^{\prime}-kk^{\prime})}{P^{\prime}M+(pp^{\prime}-kk^{\prime})}\\&\times\left(\overleftrightarrow{\partial}_{M+\frac{P^{2}}{2M}}-\overleftrightarrow{\partial}_{P^{\prime}}\right)+\frac{pP^{\prime}-p^{\prime}M}{P^{\prime}M+(pp^{\prime}-kk^{\prime})}\left(\overleftrightarrow{\partial}_{p}+\overleftrightarrow{\partial}_{p^{\prime}}\right)\\&+\frac{k^{\prime}M+kP^{\prime}}{P^{\prime}M+(pp^{\prime}-kk^{\prime})}\left(\overleftrightarrow{\partial}_{k}-\overleftrightarrow{\partial}_{k^{\prime}}\right)\Bigg]-\frac{P^{\prime}\left[P^{2}-\left(M+\frac{P^{2}}{2M}\right)^{2}\right]}{P^{\prime}\left(M+\frac{P^{2}}{2M}\right)+pp^{\prime}-kk^{\prime}}\\&\times\left(-P^{\prime}\overleftrightarrow{\partial}_{M+\frac{P^{2}}{2M}}+p^{\prime}\overleftrightarrow{\partial}_{p}-k^{\prime}\overleftrightarrow{\partial}_{k}\right)\Bigg\}\, , 
\eeas
\beas
\frac{dB_{1}^{(2,4)}}{d\omega}+\frac{dB_{1}^{(4,2)}}{d\omega}=&\frac{G}{\pi}\frac{PM^{\prime}+(pp^{\prime}-kk^{\prime})}{PM^{\prime}-(pp^{\prime}-kk^{\prime})}\Bigg[\ln\left(\frac{P\left(M^{\prime}+\frac{P^{\prime2}}{2M^{\prime}}\right)+pp^{\prime}-kk^{\prime}}{M^{\prime}P}\right)^{2}\\&+\frac{M^{\prime}}{P^{\prime}}\ln\frac{M^{\prime}-P^{\prime}}{M^{\prime}+P^{\prime}}\Bigg]\Bigg\{\left[P\left(M^{\prime}+\frac{P^{\prime2}}{2M^{\prime}}\right)+(pp^{\prime}-kk^{\prime})\right]\Bigg[\frac{PM^{\prime}-(pp^{\prime}-kk^{\prime})}{PM^{\prime}+(pp^{\prime}-kk^{\prime})}\\&\times\left(\overleftrightarrow{\partial}_{P}-\overleftrightarrow{\partial}_{M^{\prime}+\frac{P^{\prime2}}{2M^{\prime}}}\right)-\frac{p^{\prime}P-pM^{\prime}}{PM^{\prime}+(pp^{\prime}-kk^{\prime})}\left(\overleftrightarrow{\partial}_{p}+\overleftrightarrow{\partial}_{p^{\prime}}\right)\\&+\frac{(kM^{\prime}+k^{\prime}P)}{PM^{\prime}+(pp^{\prime}-kk^{\prime})}\left(\overleftrightarrow{\partial}_{k}-\overleftrightarrow{\partial}_{k^{\prime}}\right)\Bigg]+\frac{P\left[P^{\prime2}-\left(M^{\prime}+\frac{P^{\prime2}}{2M^{\prime}}\right)^{2}\right]}{P\left(M^{\prime}+\frac{P^{\prime2}}{2M^{\prime}}\right)+pp^{\prime}-kk^{\prime}}\\&\times\left(-P\overleftrightarrow{\partial}_{M^{\prime}+\frac{P^{\prime2}}{2M^{\prime}}}+p\overleftrightarrow{\partial}_{p^{\prime}}-k\overleftrightarrow{\partial}_{k^{\prime}}\right)\Bigg\}\, , 
\eeas
\beas
&\frac{dB_{1}^{(1,4)}}{d\omega}+\frac{dB_{1}^{(4,1)}}{d\omega}\\=&\frac{G}{\pi}\frac{MM^{\prime}-(pp^{\prime}-kk^{\prime})}{MM^{\prime}+(pp^{\prime}-kk^{\prime})}\Bigg[\frac{\left(M+\frac{P^{2}}{2M}\right)\left(M^{\prime}+\frac{P^{\prime2}}{2M^{\prime}}\right)-(pp^{\prime}-kk^{\prime})}{\sqrt{\left[\left(M+\frac{P^{2}}{2M}\right)\left(M^{\prime}+\frac{P^{\prime2}}{2M^{\prime}}\right)-(pp^{\prime}-kk^{\prime})\right]^{2}-M^{2}M^{\prime2}}}\\&\times\ln\frac{\left(M+\frac{P^{2}}{2M}\right)\left(M^{\prime}+\frac{P^{\prime2}}{2M^{\prime}}\right)-(pp^{\prime}-kk^{\prime})+\sqrt{\left[\left(M+\frac{P^{2}}{2M}\right)\left(M^{\prime}+\frac{P^{\prime2}}{2M^{\prime}}\right)-(pp^{\prime}-kk^{\prime})\right]^{2}-M^{2}M^{\prime2}}}{\left(M+\frac{P^{2}}{2M}\right)\left(M^{\prime}+\frac{P^{\prime2}}{2M^{\prime}}\right)-(pp^{\prime}-kk^{\prime})-\sqrt{\left[\left(M+\frac{P^{2}}{2M}\right)\left(M^{\prime}+\frac{P^{\prime2}}{2M^{\prime}}\right)-(pp^{\prime}-kk^{\prime})\right]^{2}-M^{2}M^{\prime2}}}\\&+\frac{M}{P}\ln\frac{M-P}{M+P}+\frac{M^{\prime}}{P^{\prime}}\ln\frac{M^{\prime}-P^{\prime}}{M^{\prime}+P^{\prime}}\Bigg]\Bigg\{\left[\left(M+\frac{P^{2}}{2M}\right)\left(M^{\prime}+\frac{P^{\prime2}}{2M^{\prime}}\right)-(pp^{\prime}-kk^{\prime})\right]\\&\times\Bigg[\frac{MM^{\prime}+(pp^{\prime}-kk^{\prime})}{MM^{\prime}-(pp^{\prime}-kk^{\prime})}\left(\overleftrightarrow{\partial}_{M+\frac{P^{2}}{2M}}-\overleftrightarrow{\partial}_{M^{\prime}+\frac{P^{\prime2}}{2M^{\prime}}}\right)+\frac{pM^{\prime}+p^{\prime}M}{MM^{\prime}-(pp^{\prime}-kk^{\prime})}\left(\overleftrightarrow{\partial}_{p}-\overleftrightarrow{\partial}_{p^{\prime}}\right)\\&+\frac{kM^{\prime}-k^{\prime}M}{MM^{\prime}-(pp^{\prime}-kk^{\prime})}\left(\overleftrightarrow{\partial}_{k}+\overleftrightarrow{\partial}_{k^{\prime}}\right)\Bigg]-\frac{\left(M+\frac{P^{2}}{2M}\right)\left[P^{\prime2}-\left(M^{\prime}+\frac{P^{\prime2}}{2M^{\prime}}\right)^{2}\right]}{\left(M+\frac{P^{2}}{2M}\right)\left(M^{\prime}+\frac{P^{\prime2}}{2M^{\prime}}\right)-(pp^{\prime}-kk^{\prime})}\\&\times\left[\left(M+\frac{P^{2}}{2M}\right)\overleftrightarrow{\partial}_{M^{\prime}+\frac{P^{\prime2}}{2M^{\prime}}}+p\overleftrightarrow{\partial}_{p^{\prime}}-k\overleftrightarrow{\partial}_{k^{\prime}}\right]+\frac{\left(M^{\prime}+\frac{P^{\prime2}}{2M^{\prime}}\right)\left[P^{2}-\left(M+\frac{P^{2}}{2M}\right)^{2}\right]}{\left(M+\frac{P^{2}}{2M}\right)\left(M^{\prime}+\frac{P^{\prime2}}{2M^{\prime}}\right)-(pp^{\prime}-kk^{\prime})}\\&\times\Bigg[\left(M^{\prime}+\frac{P^{\prime2}}{2M^{\prime}}\right)\overleftrightarrow{\partial}_{M+\frac{P^{2}}{2M}}+p^{\prime}\overleftrightarrow{\partial}_{p}-k^{\prime}\overleftrightarrow{\partial}_{k}\Bigg]\Bigg\}\, , 
\eeas
\beas
&\frac{dB_{1}^{(2,3)}}{d\omega}+\frac{dB_{1}^{(3,2)}}{d\omega}\\
=&\frac{2G}{\pi}\frac{PP^{\prime}-(pp^{\prime}-kk^{\prime})}{PP^{\prime}+(pp^{\prime}-kk^{\prime})}\ln\left[\frac{PP^{\prime}-(pp^{\prime}-kk^{\prime})}{2PP^{\prime}}\right]\Bigg\{\left[PP^{\prime}+(pp^{\prime}-kk^{\prime})\right]\left(\overleftrightarrow{\partial}_{P}-\overleftrightarrow{\partial}_{P^{\prime}}\right)\\&+\left(pP^{\prime}+p^{\prime}P\right)\left(\overleftrightarrow{\partial}_{p}-\overleftrightarrow{\partial}_{p^{\prime}}\right)+\left(kP^{\prime}-k^{\prime}P\right)\left(\overleftrightarrow{\partial}_{k}+\overleftrightarrow{\partial}_{k^{\prime}}\right)\Bigg\}\, . 
\eeas

\end{document}